\documentclass[showpacs,showkeys,eqsecnum,nofootinbib,reprint,author-numerical,superscriptaddress]{revtex4-1}
\usepackage{bm}
\usepackage{epsfig}
\usepackage{amsmath}
\usepackage{amsfonts}
\usepackage{amssymb}
\usepackage{color}
\usepackage{graphicx}
\usepackage{url,hyperref}
\usepackage{mathptmx}
\usepackage{epstopdf}
\usepackage{subcaption}
\usepackage[bottom]{footmisc}
\usepackage{here}
\hypersetup{
	colorlinks   = true, 
	urlcolor     = blue, 
	linkcolor    = blue, 
	citecolor   = blue 
}
\begin{document}
\title{Bohr Hamiltonian with trigonometric Pöschl-Teller potential in $\gamma-$unstable and $\gamma-$stable pictures}
\author{A.~ Ait Ben Hammou}
\affiliation{High Energy Physics and Astrophysics Laboratory, Department of Physics, Faculty of Sciences Semlalia, Cadi Ayyad University P.O.B 2390, Marrakesh 40000, Morocco.}
\author{M.~Chabab}
\affiliation{High Energy Physics and Astrophysics Laboratory, Department of Physics, Faculty of Sciences Semlalia, Cadi Ayyad University P.O.B 2390, Marrakesh 40000, Morocco.}
\author{A.~El Batoul}
\affiliation{High Energy Physics and Astrophysics Laboratory, Department of Physics, Faculty of Sciences Semlalia, Cadi Ayyad University P.O.B 2390, Marrakesh 40000, Morocco.}
\author{A.~Lahbas}
\affiliation{High Energy Physics and Astrophysics Laboratory, Department of Physics, Faculty of Sciences Semlalia, Cadi Ayyad University P.O.B 2390, Marrakesh 40000, Morocco.}
\affiliation{ESMAR, Faculty of Sciences, Mohammed V University P.O.B 1014, Rabat, Morocco.}
\author{M.~Hamzavi}
\affiliation{Department of Mathematics and Statistics, University of Texas at El Paso, El Paso, TX, USA.}
\author{I.~Moumene}
\affiliation{High Energy Physics and Astrophysics Laboratory, Department of Physics, Faculty of Sciences Semlalia, Cadi Ayyad University P.O.B 2390, Marrakesh 40000, Morocco.}
\author{M.~Oulne}
\email[Coresponding author: ]{oulne@uca.ac.ma}
\affiliation{High Energy Physics and Astrophysics Laboratory, Department of Physics, Faculty of Sciences Semlalia, Cadi Ayyad University P.O.B 2390, Marrakesh 40000, Morocco.}
\date{\today}
\begin{abstract}
\noindent
In this paper, we present an analytical solution for the Bohr Hamiltonian with the trigonometric Pöschl-Teller (P.T) potential in the cases of $\gamma$-unstable nuclei and $\gamma$-stable axially symmetric prolate deformed ones with $\gamma \approx 0$. The energy spectra and corresponding wave functions are derived by means of the asymptotic iteration method. In addition, B(E2) transition rates are calculated and compared with experimental data. Overall good agreement is obtained for inter and intra-band transitions within ground-state  and $\beta$ bands. Our numerical results, particularly for transition rates are much closer to experimental ones in comparison with those obtained by  Davidson and Kratzer potentials which are widely used in the litterature.
\end{abstract}
\maketitle
\section{Introduction}
\label{Sec1}
Like many other quantum systems, particularly molecules, atomic nuclei possess not only individual states but also so-called collective states because they involve  nucleons in rotational or vibrational motions. Indeed, besides  nuclear phenomena  characterising the individual nucleon motion, there are others such as fission processes, or large quadrupole moments, involving the collective behavior of the nucleus as a whole. The fact that quadrupole moments are observed means that the nuclei are well deformed. So, the average potential, which simulates at the lowest order the effects of all  interactions between individual nucleons,  is also deformed. Many models have been developed to describe these types of nuclei. The latest developments in the theory of collective excitations have emerged with the two important phenomenological models, namely : the interacting boson model (IBM)\cite{Iachello1} of Arima and Iachello and the geometric model of Bohr and Mottelson\cite{bohr}. The IBM exploits group theory techniques and in some cases gives analytical expressions for excitation spectra. These cases occur when the Hamiltonian has a particular symmetry, and the model does not explicitly appeal to the idea of nuclear shape. The microscopic foundations of this model are based on the shell model. Indeed, the bound states with low energy and positive parity are defined by the interaction of bosons s and d. The connection with the shell model is made by supposing that these bosons $s$ and $d$ correspond to correlated pairs of valence nucleons  coupled to $J^+=0^+$ and $J^+=2^+$. 
The corresponding symmetry of such bosons (s and d) is the U(6) mathematical group structure. It has three  solvable real dynamical symmetries U(5), SU(3), and O(6), geometrically corresponding to spherical vibration, axial symmetric rotation, and $\gamma$ unstable rotation, respectively. As a matter of fact, these dynamical symmetries are located at vertices in a diagram of what is terminologically called the Casten triangle representing the nuclear phase diagram\cite{Casten1,Cejnar1} and are considered as verifiable benchmarks for the experiment thanks to their parameter-free solutions. However, a great progress in this context was done when analytical realisation has been offered for critical points of the phase transitions from spherical vibrator to $\gamma$-unstable rotor and from spherical vibrator to axial symmetric rotor, dubbed E(5)\cite{E5} and X(5)\cite{X5}, respectively, by using  potentials as: an infinite square well for the $\beta$ variable and an oscillator one for the $\gamma$ variable. It should be noted that the energy spectrum for these two symetries are given by the zeros of a Bessel function of half-integer and irrational indices, respectively where their representatives have been experimentally identified.
\par Furthermore, the collective Bohr-Mottelson model\cite{bohr} as well as its practical implementation afforded us with an alternative description of nuclear collective excitations, which in contrast to the algebric models, is of a geometrical nature. Indeed, Bohr-Mottelson's geometric model  describes the collective excitations of nuclei in terms of surface oscillations of a drop of liquid. The resolution of its Hamiltonian in the general case requires the calculation of a potential $V(\beta,\gamma)$ which depends on the parameters $\beta$ and $\gamma$ ($\beta$ denotes the ellipsoidal deformation and $\gamma$ is the measure of axial asymmetry). This realistic theoretical model was able to describe successfully the low energy collective states and the electromagnetic transitions of a large number of even-even nuclei. The connection between this  model and the IBM comes out from considering the IBM as a second quantization of the shape variables $\beta$ and $\gamma$. 
\par So, we start by recalling the analysis of surface oscillations of nuclei (only for small deformations) given by Bohr and Mottelson. According to  their phenomenological assumptions, the nuclear surface of the deformed nucleus is an ellipsoid arbitrarily oriented in space and described, for our purposes, by a second-order deformation such as 
\begin{equation}
R(\theta,\phi)=R_0\left( 1+ \sum^{+2}_{\mu=-2}\alpha_{2,\mu}Y_{2,\mu}(\theta,\phi) \right),
\label{Eq1}
\end{equation}
where, for reality, $\alpha_{2,\mu}$ are tensors describing the deformations of the nucleus. They are expressed in terms of  the radius of the spherical nucleus $R_0$ and  the spherical harmonics $Y_{\lambda,\mu}$ as follows 
\begin{equation}
\alpha_{2,\mu}=(-1)^{\mu}\alpha^*_{2,-\mu}=\frac{1}{R_0}\int R(\theta,\phi)Y_{2,\mu}(\theta,\phi)d\Omega.
\label{Eq2}
\end{equation}
Nevertheless, for geometric reasons, it is indespensable to transform to a coordinate system which is "fixed" in the oscillating body. The collective coordinates in the body-fixed system are then linked to the space-fixed system by the following transformation 
\begin{equation}
a_{2,\mu}=\sum_{\mu}\alpha_{2,\mu}\mathcal{D}_{\mu,\nu}\left(\theta_i\right),
\label{Eq3}
\end{equation}
where the $\mathcal{D}_{\mu,\nu}\left(\theta_i\right)$ are the transformation functions (Wigner-D functions) for the spherical harmonics of second order and $\theta_i$ represents the triad of Eulerian angles $\theta$, $\varphi$, $\psi$ describing the relative orientation of the axes. Because of the symmetry of the ellipsoid about the principal axes in the proper coordinate frame, we have $a_{2,1}=a_{2,-1}=0$ and $a_{2,2}=a_{2,-2}$. Thus the five variables $a_{\lambda,\mu}$ are replaced by the three Eulerian angles $\theta_i$ and  the two real internal coordinates $a_{2,0}$ and $a_{2,2}$. Again, in place of $a_{2,0}$ and $a_{2,2}$ it is convenient to introduce the parameters $\beta$ and $\gamma$ by means of the relations : $a_{2,0}=\beta\cos\gamma$ and $a_{2,2}=a_{2,-2}=(\beta/\sqrt{2})\sin\gamma$, respecting  the conventions of D.L. Hill and J. A. Wheeler\cite{Hill}.
In terms of the parameters $\beta$ and $\gamma$, the original collective Bohr Hamiltonian in the five-dimensional form is written as \cite{bohr}:
\begin{eqnarray}
H=\frac{-\hbar^2}{2B}\Bigg[ \frac{1}{\beta^4}\frac{\partial}{\partial\beta}\beta^4\frac{\partial}{\partial \beta}+\frac{1}{\beta^2 sin3\gamma}\frac{\partial}{\partial\gamma}sin3\gamma \frac{\partial}{\partial \gamma}\nonumber\\ -\frac{1}{4\beta^2}\sum_{k}\frac{Q^2_{k}}{sin^2(\gamma-\frac{2}{3}\pi k)}\Bigg]+V(\beta, \gamma),
\label{Eq4}
\end{eqnarray}
where $B$ is the mass parameter, while $Q_{k}$ represents the angular momentum in the
variables $\theta_i$. 
Solutions of this Hamiltonian can, however, be obtained in three simple cases, according to the choosen form  for the potential  $V(\beta,\gamma)$.
\par Theoretically, the critical point symmetries have usually opened the way for the construction of other models by making use of different  phenomenological potentials leading to new exactly or quasi-exactly separable models enabling the description of nuclei which are near or far from existing critical point symmetries. Therefore, in this work, we aim to investigate the Hamiltonian (\ref{Eq4}) in the framework of  the $\gamma$-unstable  and $\gamma$-stable nuclei by considering a  potential, having a motivating physical structure, dubbed the trigonometric Pöschl-Teller potential  which takes the following form\cite{Liu,Hamzavi}:
\begin{equation}
V(\eta)=\frac{V_1}{\sin^2(\alpha \eta)}+\frac{V_2}{\cos^2(\alpha \eta)},
\label{Eq5}
\end{equation}
where the parameters $V_1$  and $V_2$  describe the property of the potential  while the parameter $\alpha$ is related to its range  where $\alpha\eta\in[0,\frac{\pi}{2}]$\cite{Liu}.
The above potential has multiple real minima only if $V_2\ne0$, and which are localized at 
\begin{equation}
\eta_0^{\pm}=\pm\arctan \left( {\frac {\sqrt [4]{V_{{1}}{V_{{2}}}^{3}}}{V_{{2}}}}
\right) {\alpha}^{-1}.
\label{Eq5a}
\end{equation}
Such a potential has been investigated in the framework of the Schr\"{o}dinger equation in \cite{Chabab}, where the bound states were well studied.
The expressions for the energy spectrum as well as the  wave functions are obtained in closed analytical form by means of the Asymptotic Iteration Method (AIM)\cite{AIM1}, an efficient technique that we have used to solve many similar problems\cite{AIM2,AIM3,AIM4,AIM5,AIM6,AIM7,AIM8,AIM9}. The present paper is organized as follows. In Section (\ref{Sec2}) the basic equations of the model are constructed, while Section (\ref{Sec3}) is devoted to the numerical calculations for energy spectra and B(E2) transition probabilities  along with their comparisons with  experimental data and  results from other models. Finally, Section (\ref{Sec4}) contains the conclusion.
\section{Basic equations of the model}
\label{Sec2}
\subsection{Trigonometric Pöschl-Teller potential in the $\gamma-$unstable case }
\begin{figure}[!h]
	\centering
	\rotatebox{0}{\includegraphics[width=80mm,height=55mm]{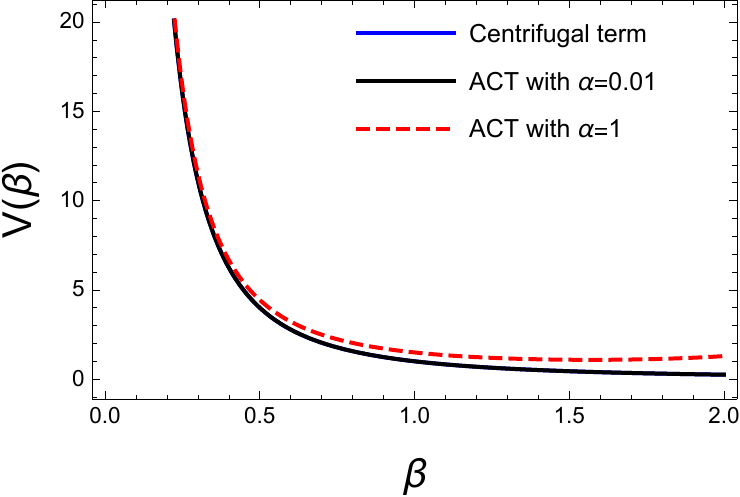}} 
	\caption{(Color online) The centrifugal term $1/\beta^2$ (blue line) and its approximations (\ref{Eq11}).}
	\label{Fig_1a}
\end{figure}

For $\gamma-$unstable structure case, the potential energy is normaly independent of $\gamma$, namely $V(\beta,\gamma)=V(\beta)$. The separation of variables in the collective Schr\"{o}dinger equation corresponding to the Hamiltonian (\ref{Eq4}) is achieved by considering a total wave function of the form \cite{wilet} :
\begin{equation}
\Psi(\beta,\gamma,\theta_i)=\xi(\beta)\Phi(\gamma,\theta_i),
\label{Eq6}
\end{equation}
where $\theta_i(i=1,2,3)$ are the Euler angles. \\
Then, the separation of the variables leads to a system of two  differential equations coupled with them by a constant $\Lambda$ :
\begin{equation}
\left[ -\frac{1}{\beta^4}\frac{\partial}{\partial\beta} {\beta^4}\frac{\partial}{\partial\beta}+u(\beta)+\frac{\Lambda}{\beta^2}\right]\xi(\beta)=\epsilon \xi(\beta),
\label{Eq7}
\end{equation}
and 
\begin{multline}
\left[- \frac{1}{\sin3\gamma}\frac{\partial}{\partial\gamma}\sin3\gamma\frac{\partial}{\partial\gamma}+
\frac{1}{4}\sum_{k}\frac{Q_{k}^{2}}{\sin^2(\gamma-\frac{2}{3}\pi k)}\right]\Phi(\gamma,\theta_i)\\=\Lambda\Phi(\gamma,\theta_i),\label{Eq8}
\end{multline}
where we have introduced the reduced energy  $\epsilon=2BE/\hbar^2$ and reduced potential $u=2BV/\hbar^2$, and the angular functions $\Phi_{\tau}(\gamma,\theta_i)$ have the form 
\begin{equation}
\Phi_{\tau}(\gamma,\theta_i)=\frac{1}{4\pi}\sqrt{\frac{(2\tau+3)!!}{\tau!}}\left( \frac{\rho_2}{\beta^2}\right)^{\tau}.
\label{Eq9}
\end{equation}
$\rho_2$ which is a function of $\beta$, $\gamma$ and $\theta_i$, will be introduced later.
Here we mention that the first wave equation contains only the $\beta$ variable, while the second contains the $\gamma$ variable and the Euler angles. The $\gamma$ and Euler angles equation \eqref{Eq8} has been solved by B\`es\cite{bes}. In this equation, the eigenvalues of the second-order Casimir operator SO(5) are expressed in the following form $\Lambda= \tau(\tau + 3)$, where $\tau$ is the seniority quantum number, characterizing the irreducible representations of SO(5) and taking the values $\tau=0, 1, 2, ...$ \cite{rakavy}. The values of angular momentum $L$ occurring for each $\tau$ are provided by a well known algorithm and are listed in \cite{Iachello1}. The ground state band levels are determined by $L=2\tau$.
By inserting the proposed potential \eqref{Eq5} in the radial equation \eqref{Eq7} and using the transformation of the wave function $\xi(\beta)=f(\beta)/\beta^2$, we easily obtain :
\begin{equation}
\left[ -\frac{\partial^2}{\partial\beta^2}-\left(\frac{V_1}{\sin^2(\alpha \beta)}+\frac{V_2}{\cos^2(\alpha \beta)}\right)+\frac{\Lambda+2}{\beta^{2}}\right] f(\beta)=\epsilon f(\beta). 
\label{Eq10}
\end{equation}
For $L$-wave functions, the above equation cannot be solved analytically. Then, in order to obtain quasi-analytical solutions, we need to employ an approximation to the centrifugal term (ACT) as\cite{Hamzavi} 
\begin{equation}
\frac{1}{\beta^2}=\lim\limits_{\alpha \rightarrow 0} \alpha^2\left(d_{0}+\frac{1}{sin^2(\alpha\beta)}\right),
\label{Eq11}
\end{equation}
where $ d_{0}=\frac{1}{12}$ is a dimensionless shifting parameter.
In order to show the validity and accuracy of our choice to the above approximation scheme, we  plot the centrifugal potential term $1/\beta^2$ and the approximation : $\alpha^2\left(d_{0}+\frac{1}{sin^2(\alpha\beta)}\right)$ with $\alpha=0.01$ and $\alpha=1$ as a function of $\beta$ in Figure (\ref{Fig_1a}). As clearly illustrated, the three curves coincide together and show how accurate is this replacement especially in the vicinity of $\alpha=0$. 
\par Therefore, by using the new variable $y=sin^2(\alpha \beta)$, Eq. (\ref{Eq10}) reduces to 
\begin{equation}
f''(y)+\frac{2 y-1}{2 y(y-1)} f'(y)+\frac{(A_1y^2+A_2y-A_3)}{4  \alpha^2 y^2(y-1)^2} f(y),
\label{Eq12}
\end{equation}
with :
\begin{align}
&A_1=d(2+\Lambda)\alpha^2-\epsilon,\nonumber\\
&A_2=-(2+\Lambda)(d-1) \alpha^2+\epsilon+V_1-V_2,\\
&A_3=(\Lambda+2)\alpha^2+V_1\nonumber.
\label{Eq13}
\end{align}
In order to apply AIM, we propose the  following  ansatz :
\begin{equation}
f(y)=y^{\mu} (1-y)^{\nu} \phi(y).
\label{Eq14}
\end{equation}
Using the  AIM, we can obtain the generalized formula of the energy spectrum  in the following  form :
\begin{equation}
\epsilon^{(\gamma-uns)}=\frac{\alpha^2}{24}\left(\frac{L(L+1)}{3}+2\right)+2 \alpha^2 (n_{\beta}+\mu+\nu)^2,
\label{Eq15}
\end{equation}
where 
\begin{equation}
\mu=\dfrac{1}{4}+\dfrac{1}{4}\sqrt{4 \Lambda+9+\dfrac{4V_1}{\alpha^2}},\ \nu=\dfrac{1}{4}+\dfrac{1}{4}\sqrt{1+\dfrac{4V_2}{\alpha^2}},
\label{Eq16}
\end{equation}
and $n_{\beta}$ is the principal quantum number.
The corresponding wave function $\xi(y)$ for $\gamma$-unstable case is written in terms of Hypergeometric function as :
\begin{equation}
\xi(y)= Ny^{ \mu} (1-y)^{\nu} _{2}F_{1}[-n,2\mu+2\nu+n,2 \mu+\frac{1}{2},y],
\label{Eq17}
\end{equation}
where $N$ is a normalization constant 
\begin{equation}
N=\left[\frac{2 \alpha(2\mu +2\nu+2n)\Gamma[2 \mu+2\nu+n]\Gamma[2 \mu+\frac{1}{2}+n]}{n!\Gamma[2 \mu+\frac{1}{2}]^2 \Gamma[2 \nu+\frac{1}{2}+n]}\right]^{1/2}.
\label{Eq18}
\end{equation}
\subsection{Trigonometric Pöschl-Teller potential in the $\gamma-$stable case }
In this case, the exact separation of the collective variables $\beta$ and $\gamma$ in Eq. (\ref{Eq4}) may be achieved by considering the following expression for the potential energy :
\begin{equation}
V(\beta,\gamma)=V(\beta)+W(\gamma)/\beta^2,
\label{Eq19}
\end{equation}
The treatment of the $\gamma$ degree of freedom is combined with  the Pöschl-Teller potential. So, for the $\gamma$-part, we use a harmonic oscillator potential
\begin{equation}
 W(\gamma)=(3c)^2\gamma^2.
 \label{Eq20}
\end{equation}
In this step, we consider the total wave function of the form \cite{X5}
\begin{equation}
\Psi(\beta,\gamma,\theta_i)=F_L(\beta)\eta_K(\gamma)\mathcal{D}_{M,K}^L(\theta_i),
\label{Eq21}
\end{equation}
$L$  is the  total angular momentum, where  $M$ and $K$ are the eigenvalues of the projections of angular momentum on the laboratory fixed $x$-axis and the body-fixed $x'$-axis respectively. 
Then separation of variables leads to
\begin{equation}
\Big[ -\frac{1}{\beta^4}\frac{\partial}{\partial\beta} {\beta^4}\frac{\partial}{\partial\beta}+\frac{\hat{\Lambda}+\frac{L(L+1)}{3}}{\beta^2}+u(\beta)\Big]F_L(\beta)=\epsilon F_L(\beta),  \label{Eq22}
\end{equation}
\begin{multline}
\Big[- \frac{1}{\sin3\gamma}\frac{\partial}{\partial\gamma}\sin3\gamma\frac{\partial}{\partial\gamma}+
\frac{K^2}{4}\Big(\frac{1}{\sin^2\gamma}-\frac{4}{3}\Big) +w(\gamma) \Big]\eta_K(\gamma)\\={\hat\Lambda}\eta_K (\gamma). \label{Eq23}
\end{multline}
By solving the $\gamma$-vibrational part of the Schr\"{o}dinger equation following method of \cite{AIM2,davison}, the $\gamma$ angular wave functions can be written as \cite{X5,AIM2,davison}
\begin{equation}
\eta_{n_{\gamma},K}=N_{n_{\gamma},K}\ \gamma^{|K/2|}\ e^{-\frac{3c\gamma^2}{2}}L_{\tilde n_{\gamma}}^{|K/2|}\Big(3c\gamma^2\Big),
\label{Eq24}
\end{equation}
with $\tilde n_{\gamma}=\frac{n_{\gamma}-|K/2|}{2}$ where $n_{\gamma}$ is the quantum number related to $\gamma$ oscillations, while $L_{\tilde n_{\gamma}}^{K/2}$ represents the Laguerre polynomial and $N_{\gamma,K}$ the normalization constant,
\begin{equation}
N_{n_{\gamma},K}=\left[\frac{2}{3}(3c)^{1+|K/2|}\frac{\tilde n_{\gamma}!}{\Gamma(\tilde n_{\gamma}+|K/2|+1)}\right]^{1/2}.
\label{Eq25}
\end{equation}
By analogy with the previous case, namely the $\gamma$-unstable, the energy spectrum for the $\gamma$-stable one is obtained as follows
\begin{equation}
\epsilon^{(\gamma-sta)}=\frac{\alpha^2}{24}\left(\frac{L(L+1)}{3}+\Lambda'+2\right)+2 \alpha^2 (n_{\beta}+\mu+\nu)^2,
\label{Eq26}
\end{equation}
where 
\begin{align}
&\mu=\dfrac{1}{4}+\dfrac{1}{4}\sqrt{4\left(\frac{L(L+1)}{3}+ \Lambda' \right) +9+\dfrac{4 V_1}{\alpha^2}},\nonumber\\
&\Lambda'=6 c (n_{\gamma}+1)-  \dfrac{K^2}{3},\label{Eq27}\\
&\nu=\dfrac{1}{4}+\dfrac{1}{4}\sqrt{1+\dfrac{4 V_2}{\alpha^2}}.\nonumber
\end{align}
We emphasize here that the shape of the wave function is similar to that of the equation (\ref{Eq17}) but with the pramameters listed above.
\subsection{B(E2) Transition rates}
Electromagnetic properties of collective states are dominated by the B(E2) transition probabilities. Having the analytical expression of the obtained total wave function, one can readily compute the $B(E2)$ transition rates. In the general case the quadrupole operator is defined as \cite{wilet,IBudaca}
\begin{align}
T_{M}^{(E2)}=t\rho_2=t\beta\Bigg[\mathcal{D}^{(2)}_{M,0}(\theta_i)\cos\gamma  +\frac{1}{\sqrt{2}}\Big( \mathcal{D}^{(2)}_{M,2}(\theta_i)\nonumber\\
+\mathcal{D}^{(2)}_{M,-2}(\theta_i) \Big)\sin\gamma \Bigg],
\label{Eq28}
\end{align}
where $t$ is a scale factor.
For the $\gamma$-unstable nuclei, $B(E2)$ transition rates from an initial to a final state are given by \cite{Edmonds}
\begin{align}
B(E2;s_i,L_i  \rightarrow s_f,L_f)  =\frac{5}{16\pi} \frac{\mid \left<s_f,L_f\mid\mid T^{(E2)} \mid\mid s_i,L_i\right>\mid^2}{(2L_i+1)}\nonumber\\
=\frac{2L_f+1}{2L_i+1}B(E2;s_f,L_f  \rightarrow s_i,L_i),
\label{Eq29}
\end{align}
where $s$ denotes quantum numbers other than the angular momentum $L$.
Then, the full symmetrized wavefunction is written from equation \eqref{Eq6} as
\begin{equation}
\Psi(\beta,\gamma,\theta_i)=\beta^{-2}\chi_{n,\tau}(\beta)\Phi_{\tau}(\gamma,\theta_i).
\label{Eq30}
\end{equation}
The radial function $\chi(\beta)$ is given by Eq. \eqref{Eq17}, while the angular functions $\Phi_{\tau}(\gamma,\theta_i)$ have the form of equation \eqref{Eq9}. From Eqs. \eqref{Eq29} and \eqref{Eq30} one obtains \cite{Bonatsos}
\begin{equation}
B(E2;L_{n,\tau}\rightarrow(L+2)_{n',\tau+1})=\frac{(\tau+1)(4\tau+5)}{(2\tau+5)(4\tau+1)}t^2I_{n',\tau+1;n,\tau}^2,
\label{Eq31}
\end{equation}
with
\begin{equation}
I_{n',\tau+1;n,\tau}= \int_0^{\infty} \beta \xi_{n',\tau+1}(\beta)\xi_{n,\tau}(\beta)\beta^4d\beta.
\label{Eq32}
\end{equation}
In the other case, i.e. for $\gamma$-stable axially deformed nuclei around $\gamma=0$,  $B(E2)$ transition rates read \cite{bijker03}
\begin{align}
B(E2;nLn_{\gamma}K\longrightarrow n'L'n'_{\gamma}K')
=\frac{5}{16\pi}t^2\langle L,K,2,K'-K|L',K'\rangle^2\nonumber \\ \times I^2_{n,L;n',L'}C^2_{n_{\gamma},K;n'_{\gamma},K'},
\label{Eq33}
\end{align}
with
\begin{align}
I_{n,L;n',L'}=\int_0^{\infty} \beta \xi_{n',L'}(\beta)\xi_{n,L}(\beta)\beta^4d\beta,
\label{Eq34}
\end{align}
$C_{_{n_{\gamma}K,n'_{\gamma}K'}}$ contains the integral over $\gamma$. For $\Delta K=0$ corresponding to transitions  ($g\rightarrow g, \gamma\rightarrow\gamma, \beta\rightarrow\beta$ and $\beta\rightarrow g $), the $\gamma$-integral part reduces to the orthonormality condition of the $\gamma$-wave functions : $C_{_{n_{\gamma},K;n'_{\gamma},K'}}=\delta_{_{n_{\gamma},n'_{\gamma}}}\delta_{_{K,K'}}$.
While for $\Delta K=2$ corresponding to transitions ($\gamma\rightarrow g, \gamma\rightarrow\beta$), this
integral takes the form.
\begin{align}
C_{n_{\gamma},K;n'_{\gamma},K'}=\int^{\pi/3}_0 \sin\gamma\eta_{n_{\gamma}K}\eta_{n'_{\gamma}K'}|\sin3\gamma|d\gamma.
\label{Eq35}
\end{align}
Using the approximation $|\sin3\gamma| \approx 3|\gamma|$ and  Eq. \eqref{Eq35} the last integral becomes
\begin{align}
C_{n_{\gamma},K;n'_{\gamma},K'}=\frac{2 (3c)^{1+|K|/4+|K'|/4}}{\left(\Gamma(|K/2|+1)\Gamma(|K'/2|+1) \right)^{1/2}}\nonumber\\ \times \int^{\pi/3}_0 \gamma^{2+\frac{|K'|+|K|}{2}} e^{-3c\gamma^2} d\gamma,
\label{37}
\end{align}
where the Laguerre polynomials are unity since $\tilde n_{\gamma}=0$.
\section{NUMERICAL RESULTS}
\label{Sec3}
\begin{figure}[!h]
	\centering
	\rotatebox{0}{\includegraphics[width=80mm,height=55mm]{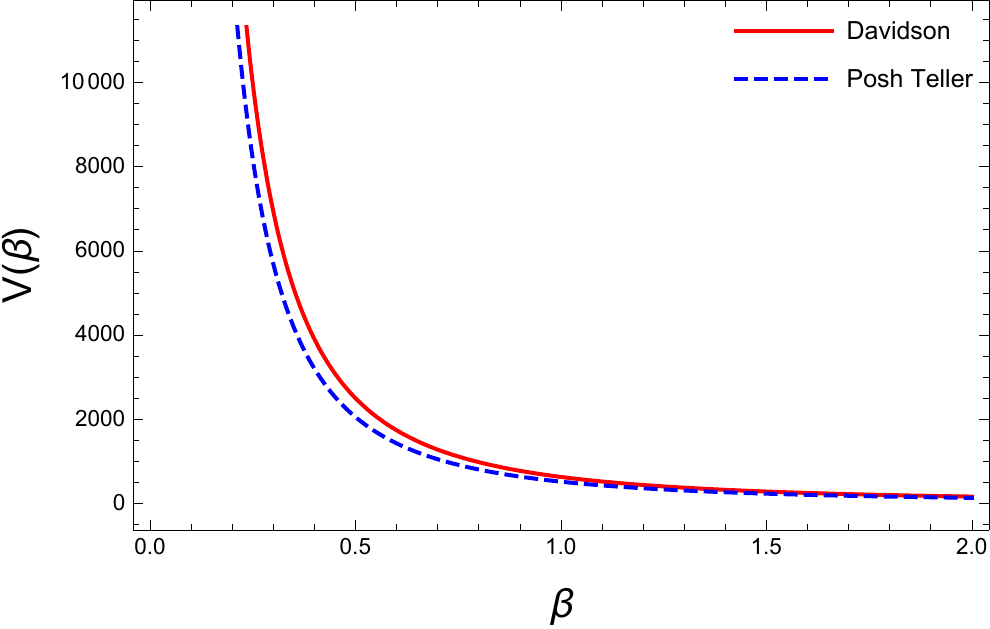}} 
	\caption{(Color online) Evolution of P.T and Davidson potentials as function of collective parameter $\beta$. The quantities shown are dimensionless. }
	\label{Fig_1}
\end{figure}

\begin{figure}[!h]
	\centering
	\rotatebox{0}{\includegraphics[width=40mm,height=50mm]{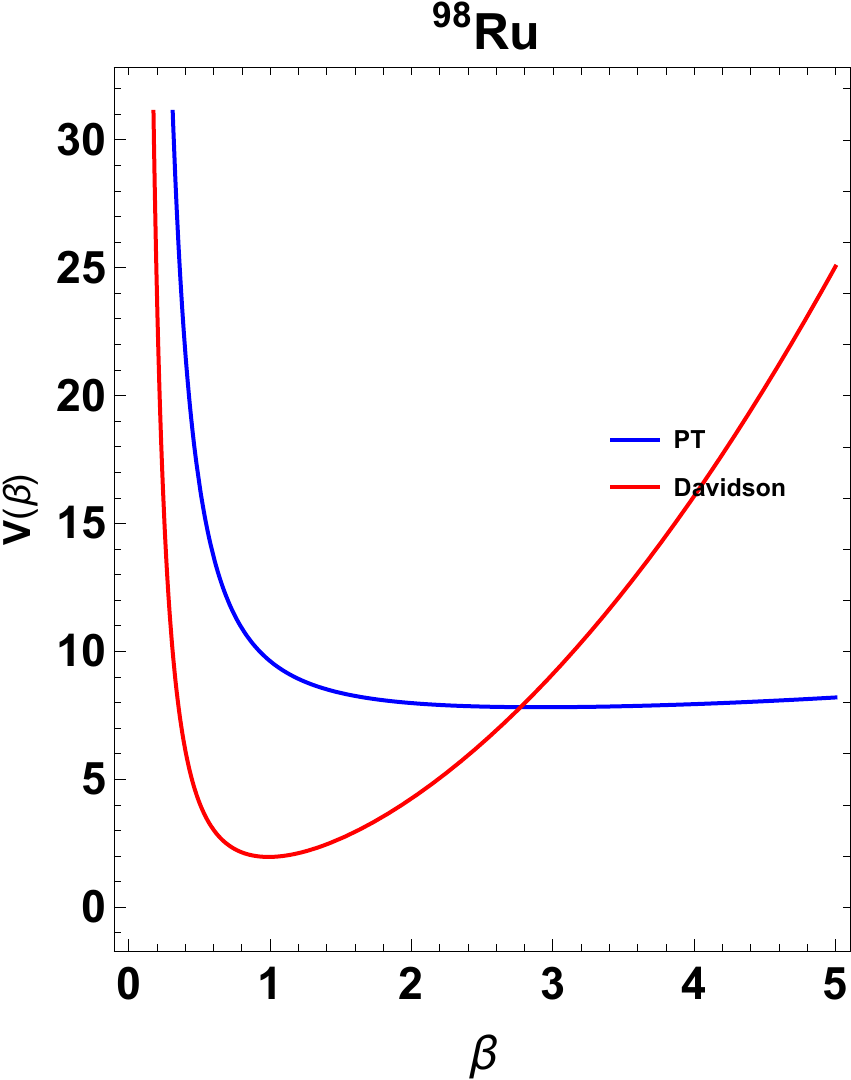}} 
	\rotatebox{0}{\includegraphics[width=40mm,height=50mm]{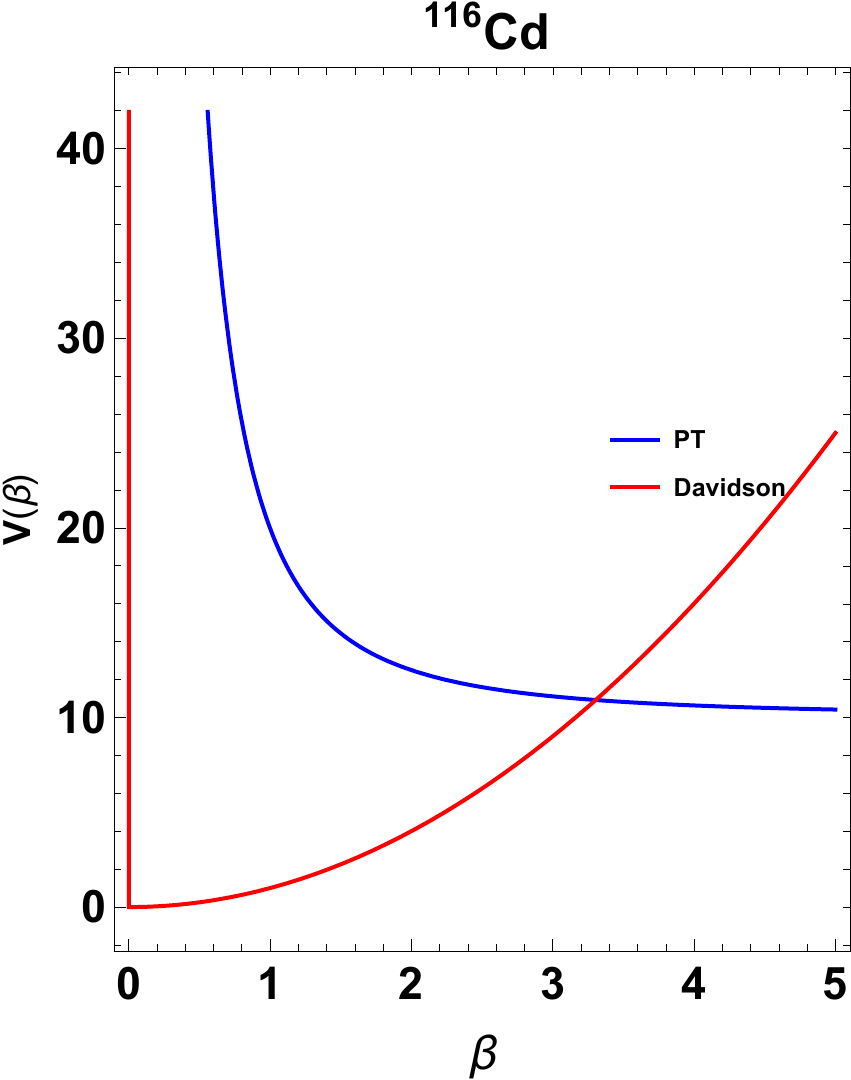}}
	\caption{(Color online) Typical evolution  of P.T and Davidson potentials for the nuclei : ${}^{98}$Ru and ${}^{116}$Cd.}
	\label{Fig_2}
\end{figure}

\begin{figure}[!h]
	\centering
	\includegraphics[width=50mm]{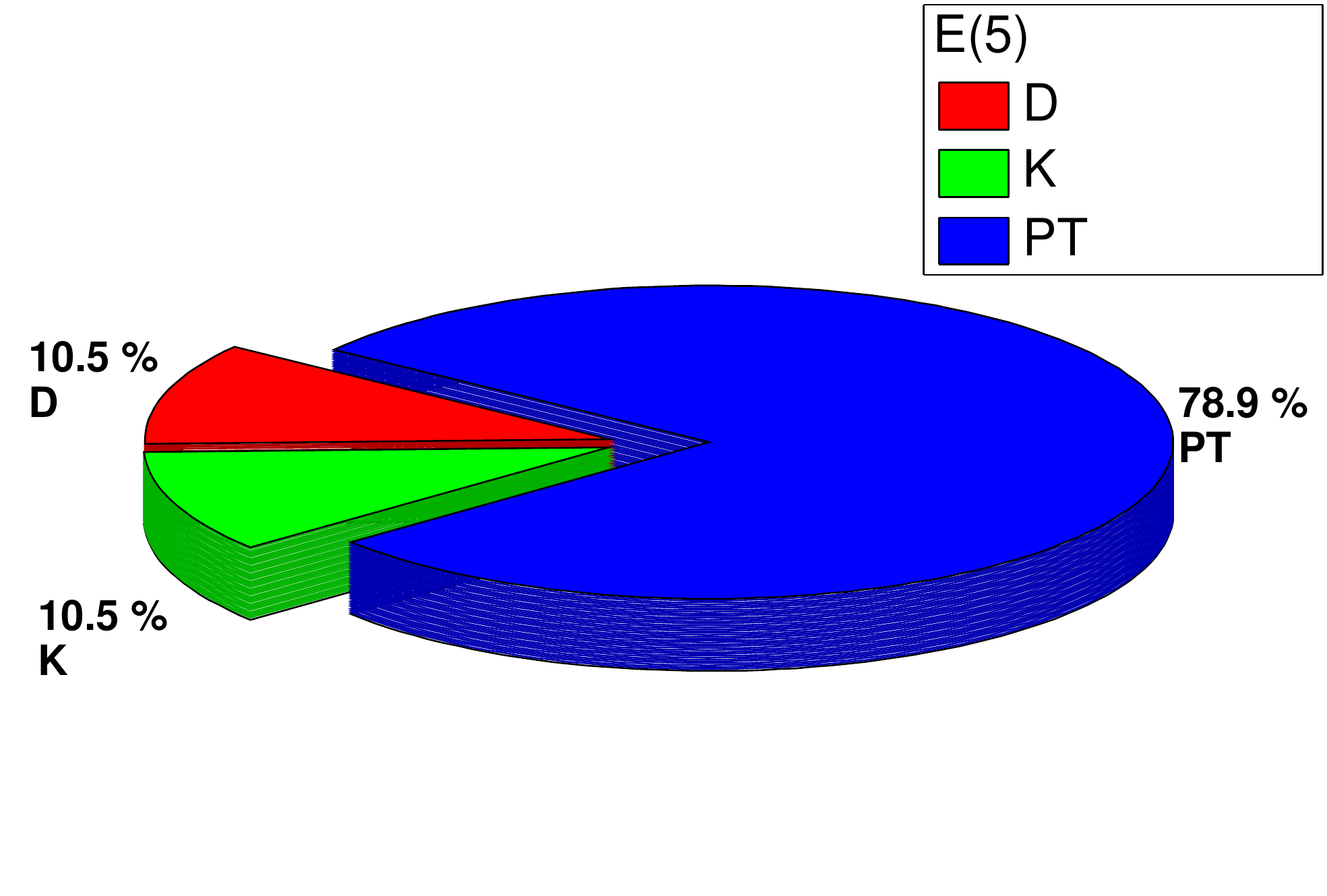}
	\includegraphics[width=50mm]{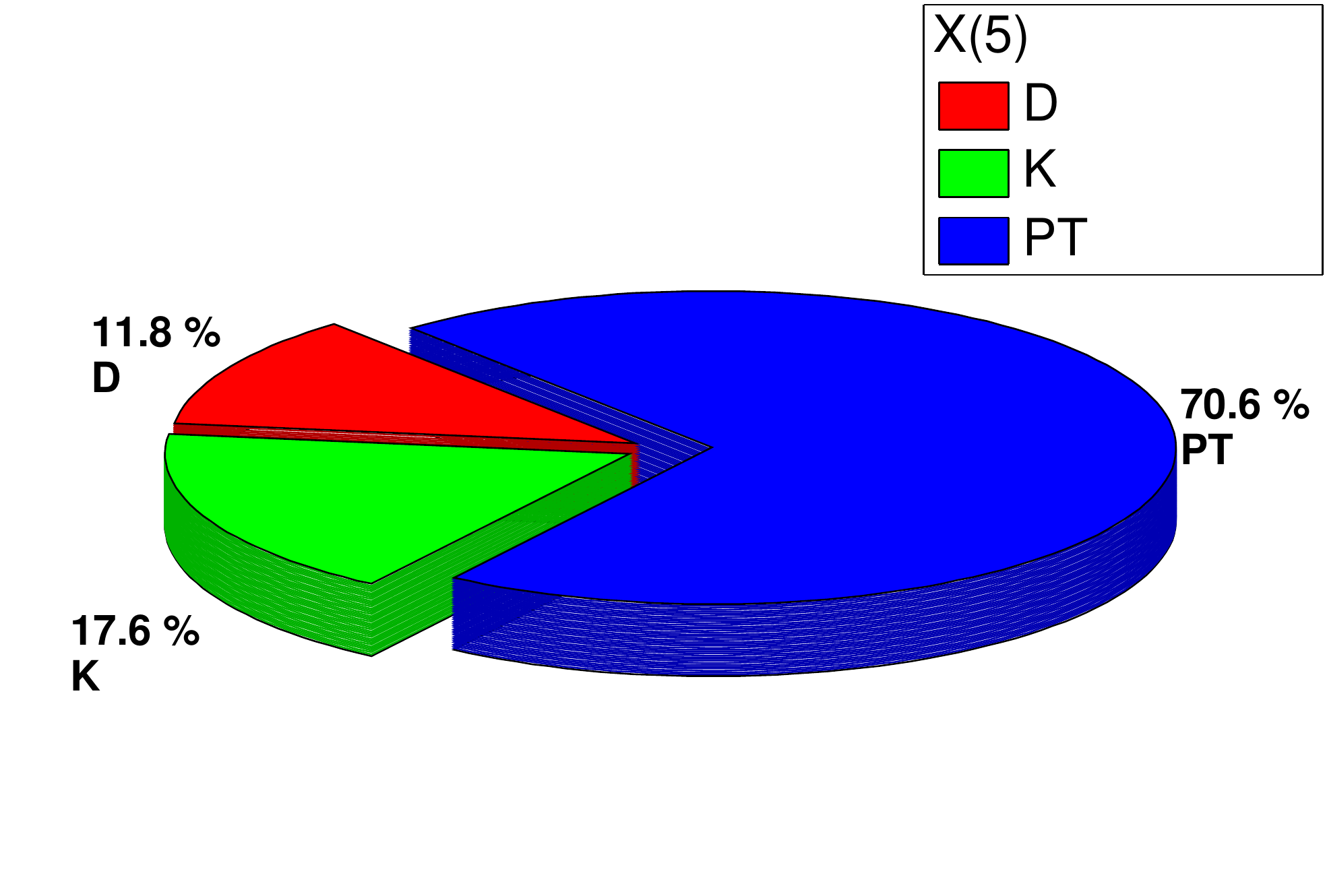}
	\caption{(Color online) The percentage of the best value of the rms in Tables (\ref{Table5}-\ref{Table6}) of $B(E2)$  transition rate obtained by P.T, Davidson and Kratzer potentials, in the $\gamma$-unstable  (top) and the $\gamma$-stable case (down).}
	\label{Fig_3}
\end{figure}

\begin{figure}[H]
	\centering
	\rotatebox{0}{\includegraphics[width=80mm,height=50mm]{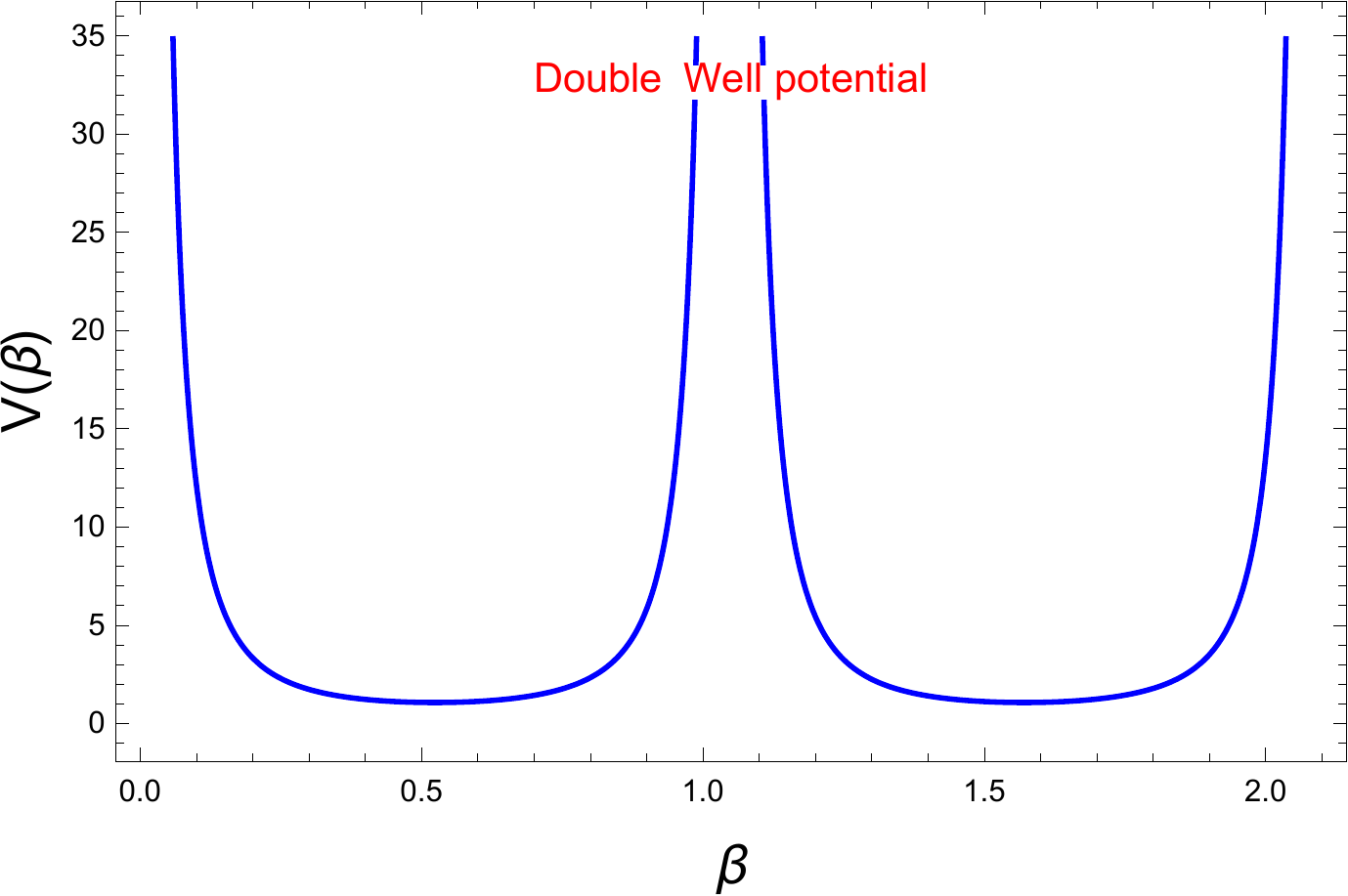}} 
	\caption{(Color online) Shape of the P. T potential showing  two minima due to the periodicity of the trigonometric functions.}
	\label{Fig_4}
\end{figure}

\begin{figure}[h]
	\centering
	\rotatebox{0}{\includegraphics[width=80mm,height=50mm]{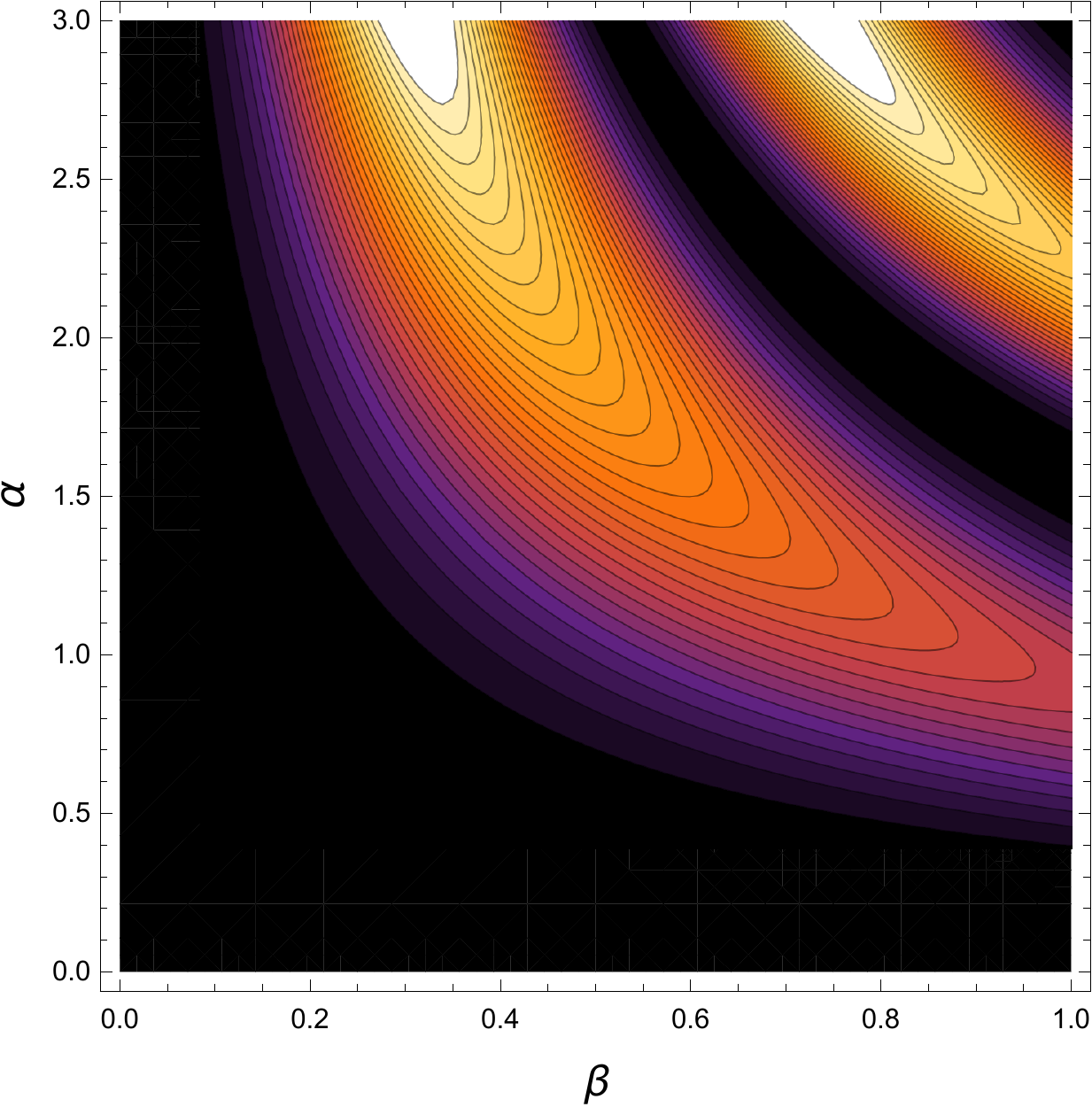}} 
	\rotatebox{0}{\includegraphics[width=80mm,height=50mm]{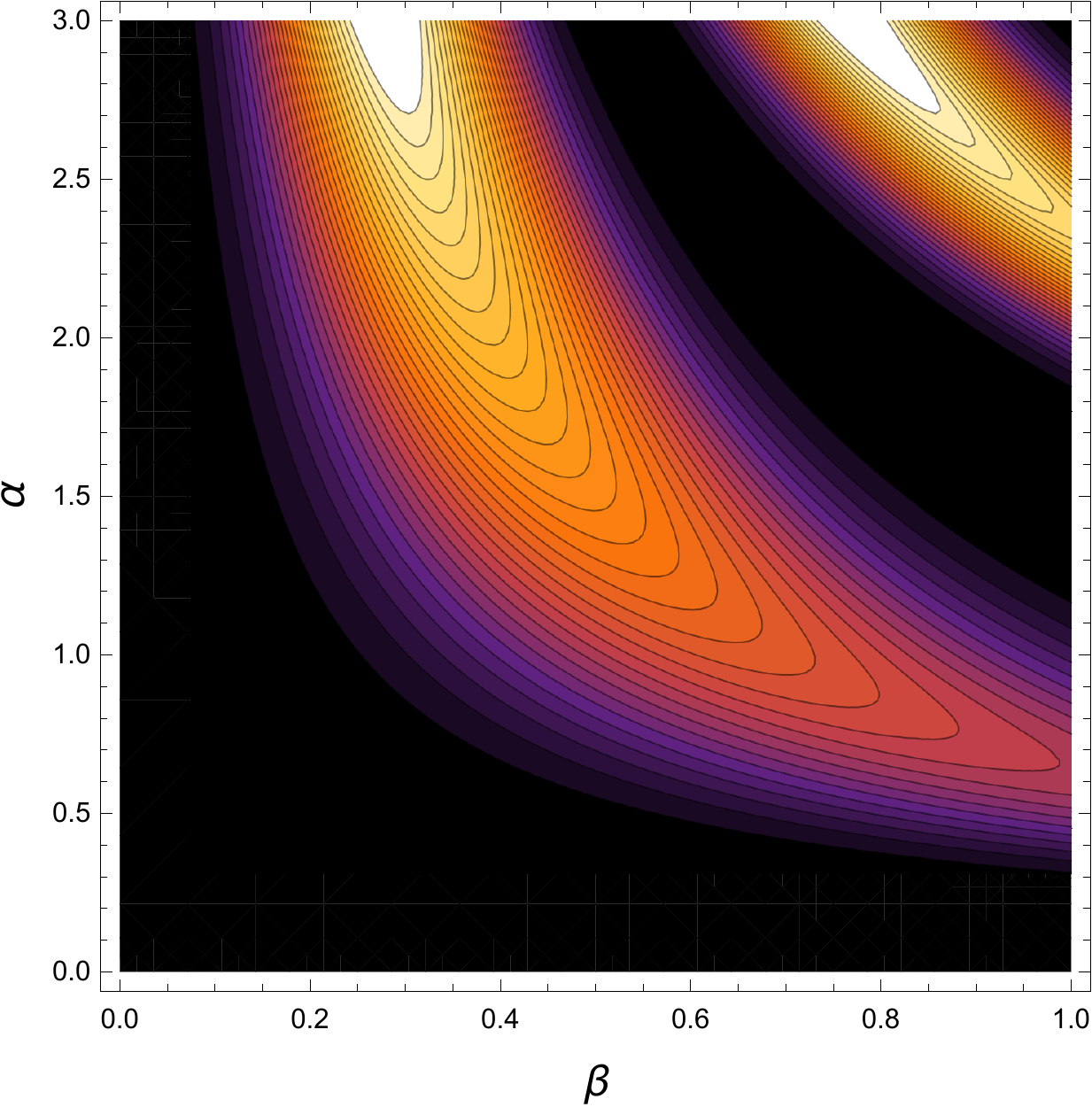}}
	\caption{(Color online) Lines of constant probability density of ground-state drawn as functions of $\alpha$ and $\beta$ show also the periodical behaviour which is previously observed in Fig. \ref{Fig_4}. The values of the parameters $V_1$ and $V_2$ used in this figure are :  $V_2=2V_1=1$ (left) and $V_2=10V_1=10$ (right).}
	\label{Fig_5}
\end{figure}

In this work, we proceed to a systematic comparison between Pöschl-Teller potential and the widely used Davidson potential in the calculation of energy spectra and transition probabilities for several even-even  transitional nuclei situated in the vicinity  of the critical point symmetries E(5) and X(5). The idea of such a comparative study issued from the observation that both potentials behave similarly for given arbitrary potential parameters as can be seen from Fig. \ref{Fig_1}. For concrete nuclei, such parameters take definite values given in Table (\ref{Table1}) and Table (\ref{Table2}) which are obtained by fitting the energy formulas in Eq. (\ref{Eq15}) and Eq. (\ref{Eq26}) on the available experimental data\cite{data}. The quality of the fit is evaluated by making use of the r.m.s deviation of the theoretical results from the experimental ones :  
\begin{equation}
\sigma_{r.m.s.}=\sqrt{\frac{\sum_{i=1}^N(E_i(exp)-E_i(th))^2}{(N-1)E(2_1^+)^2}},
\label{Eq36}
\end{equation}
where  $E_i(th)$ is the calculated energy of the $i-th$ level, $E_i(exp)$  the corresponding experimental one and $N$ the number of the considered levels. The theoretical predictions are done with Eq. (\ref{Eq15}) and Eq. (\ref{Eq26}) for low-lying bands which are classified by the principal quantum number $n$. The ground state band (g.s.) with $n=0$, the $\beta$-band with $n=1$ and the $\gamma$-band is obtained from degeneracies of the g.s. levels.
In Tables (\ref{Table3}-\ref{Table4}), we presented the obtained results for the energy within P.T potential in $\gamma$-unstable picture at the CPS E(5). In these tables, are presented also the experimental data and previousely calculated energies with Davidson potential by other authors\cite{Bonatsos2} as well as the obtained results with Kratzer potential in Ref\cite{Bonatsos3} for a supplementary comparison.
Here, we have to notice that Davidson and Kratzer results in Ref\cite{Bonatsos2} and Ref\cite{Bonatsos3} respectively, have been obtained within the Deformation Dependent Mass formalism (DDM). Furthermore, for example, in the  $\gamma$-unstable case, the DDM-Davidson uses only two fitting parameters in total, namely $\beta$ that indicates the position of the minimum of the potential and the deformation parameter $a$, while  our approach uses three parameters. Similarly, in the prolate deformed case, the DDM-Davidson in \cite{Bonatsos2} uses three parameters in total, while our model uses four parameters.
Such a formalism which introduced a further fitted parameter had improved even more the results which would have been obtained with a pure Davidson potential.
From these tables one can see that our results are comparable to those of Davidson potential and both are in a good agreement with the experiment. Such a concordance between P.T and Davidson potentials is  essentially due to the fact that both potentials have similar behavior for deformations below the potential minimum as can be seen from figure (\ref{Fig_2}).
From Table (\ref{Table3}) it comes out that the nuclei ${}^{98}$Ru, ${}^{108}$Pd, ${}^{114}$Cd, ${}^{134}$Xe and ${}^{142}$Gd are good candidate for our model but the best ones are  ${}^{98}$Ru, ${}^{134}$Xe and ${}^{142}$Gd for which $\sigma_{r.m.s.}$ is so lower and the value of the energy ratio $R_{4/2}$ is equal or very close to the experimental one, which is a characteristic of the considered CPS. A similar situation is observed from Table (\ref{Table4})  for P.T-X(5). The obtained good candidate nuclei in this case are ${}^{160}$Gd, ${}^{162}$Gd, ${}^{164}$Dy, ${}^{166}$Dy, ${}^{172}$Yb, ${}^{174}$Yb, ${}^{184}$W and ${}^{188}$Os, and the best ones are ${}^{162}$Gd, ${}^{164}$Dy, ${}^{166}$Dy and ${}^{174}$Yb. With the obtained optimal parameters in Table (\ref{Table1}) and Table (\ref{Table2}), we have calculated  intra and inter bands transition rates for several nuclei. The obtained results are presented in Table (\ref{Table5}) and Table (\ref{Table6}) compared to those obtained with Davidson and Kratzer potentials as well as to the available experimental data. From these tables, one can see clearly that the calculated transition rates with P.T potential are fairly better than those  corresponding to the Davidson and Kratzer ones. 
Over the $27$ studied nuclei, $78.9\%$ of them are well reproduced with P.T potential versus  $10.5\%$ for Davidson potential and $10.5\%$ for Kratzer one in the case of the $\gamma$-unstable CPS as it is presented in Fig. \ref{Fig_3}. 
The same situation is observed within X(5) with $70.6\%$ of nuclei are well reproduced with P.T potential versus $11.8\%$ for Davidson and $17.6\%$ for Kratzer one (for $24$ studied nuclei). Such a discrepancy between these three potentials is related to the behavior of each potential beyond its minimum. As the potential is more flat for higher values of the $\beta$ collective coordinate as the calculated transition rates are more accurate. The $\beta$ band is particularly more sensitive to the flatness of the potential as can be seen from Table (\ref{Table5}) and Table (\ref{Table6}). Such a fact corroborates the constatation we have made in our previous works \cite{AIM5,AIM7} with other potentials.
Besides the manifested performance by P.T potential versus Davidson one in reproducing the experimental data for energy spectra and particularly transition rates, it presents another advantage in respect to this latter, namely: its ability to show two minima due to the periodicity of the trigonometric functions entering in its expression. In Fig. \ref{Fig_4}, the plot of P.T potential shows a double well potential. Such a characteristic could be useful for studying shape coexistence in nuclei that will be subject of our future work.
Indeed, the study of the distribution of the probability density makes it possible to better understand the phenomenology of the collective conditions associated with such a model. The lines evolution of constant probability density as a function of the collective variable $\beta$ as well as the parameter $\alpha$ of the potential is depicted in Fig. \ref{Fig_5} for the ground state. Basically the profile of ground state probability shows a sharp and symmetrical peak centered in the deformed minimum of the potential as can be seen on the corresponding lines  constant.  Such a deformed minimum can be calculated by using  equation (\ref{Eq5a}).
On the other hand, it is crucial to note that in spite of the presence of two minima caused by the periodicity of the trigonometric functions, the ground state and its rotation excitations favor the deformed minimum which is in fact due to centrifugal stretching. On the contrary, vibratory states prefer the spherical minimum particularly in the region of lower $\beta$ deformation for E(5) limit.
\section{Conclusion}
\label{Sec4}
In this paper, the Schr\"{o}dinger equation with Bohr Hamiltonian has been analytically solved with the trigonometric Pöschl-Teller in the $\gamma$-unstable and $\gamma$-stable  regimes. Having the energy spectra formulas as well as the corresponding wave functions, we calculated energy levels normalized to the energy of the first excited state in the ground state band and $B(E2)$ for inter and intra-band transition rates for several nuclei. The comparison between our numerical results and the experimental ones has shown a good agreement and a better performance in respect to the Davidson and Kratzer potentials particularly for transition probabilities due to the flatness of Pöschl-Teller potential beyond its minimum. The fact that the Pöschl-Teller potential possesses at least two minima, as it was shown in the present work, presents an important opportunity to be used in studies of shape coexistence in nuclei within the Bohr-Mottelson model for which an appropriate double-well potential is seldom available.  	
\begin{table*}[b]
	\centering
	{\renewcommand{\arraystretch}{1}
		{\setlength{\tabcolsep}{1cm} 
			\caption{  The values of free parameters fitted to the experimental data for the $\gamma$-unstable nuclei.}
			\label{Table1}
			\begin{tabular}{|cccc|}
				\hline
				Nucleus \qquad&$V_1$&$\alpha$&$V_2$\\
				\hline
				{$^{98}$Ru}&0.009& 0.065&	7.26\\
				{$^{102}$Pd} &0.46& 0.83&6.15\\
				{$^{104}$Pd} & 0.45&0.71&6.16\\
				{$^{106}$Pd} & 0.14& 0.34&7.04\\
				{$^{108}$Pd}& 0.13& 0.82&1.52\\	
				{$^{112}$Pd}& 0.99& 0.72&0.01\\
				{$^{114}$Pd} & 0.99&0.67&0.01\\	
				{$^{106}$Cd}& 0.15& 0.19&6.90\\
				{$^{108}$Cd}&0.04& 0.14&7.07\\
				{$^{110}$Cd}&0.001&0.33&9.98\\
				{$^{114}$Cd}&0.001& 0.17&9.99\\	
				{$^{116}$Cd}& 0.001&0.017&9.99\\
				{$^{120}$Cd}& 0.001& 0.01&9.99\\
				{$^{128}$Xe}& 0.711&0.33&0.49\\
				{$^{130}$Xe}& 0.001& 0.012&9.99\\	
				{$^{132}$Xe}& 0.01& 0.05&9.968\\
				{$^{134}$Xe}&0.001& 0.07&9.99\\
				{$^{132}$Ba}& 0.13& 0.128&0.71\\
				{$^{134}$Ba}& 0.001&0.07&9.99\\
				{$^{136}$Ba}& 0.001& 0.08&9.99\\
				{$^{136}$Ce}& 0.01&0.25&	9.99\\
				{$^{140}$Nd}& 0.001& 0.14&9.99\\	
				{$^{142}$Gd}& 0.03& 0.09&7.13\\	
				{$^{144}$Gd} &0.61& 0.69&0.62\\
				{$^{194}$Pt} & 0.74& 0.39&0.03\\
				{$^{196}$Pt}& 0.99& 0.51&0.01\\
				{$^{198}$Pt}& 0.00001&0.99&	0.01\\
				\hline
	\end{tabular}}}
\end{table*}
\begin{table*}[tbph]
	\centering
	{\renewcommand{\arraystretch}{1}
		{\setlength{\tabcolsep}{1cm} 
			\caption{The values of free parameters fitted to the experimental data for the $\gamma$-stable nuclei.
			}
			\label{Table2}
			\begin{tabular}{|ccccc|}
				\hline
				Nucleus \qquad&$V_1$&$V_2$&$c$&$\alpha$\\
				\hline
				{$^{154}$Sm} & 0.001& 0.99& 6.88&0.04\\
				{$^{158}$Gd} &0.79& 8.86&	5.36& 0.24 \\
				{$^{160}$Gd} & 0.68&2.26&4.47&0.13\\
				{$^{162}$Gd}  & 0.50&5.88&4.07&0.08\\
				{$^{158}$Dy} & 0.001& 9.99&3.71& 0.13\\
				{$^{162}$Dy} & 0.64& 3.73&3.80& 0.14 \\
				{$^{164}$Dy} & 0.001& 9.99&	3.42& 0.003 \\
				{$^{166}$Dy}   & 0.70& 0.40&3.75&0.17 \\
				{$^{160}$Er} & 0.001& 9.99&2.41& 0.05 \\
				{$^{162}$Er} & 0.01&7.14&3.71& 0.08 \\
				{$^{166}$Er} & 0.001&9.99&3.44&0.004 \\
				{$^{168}$Er}  &0.70& 0.35&3.44&0.18 \\
				{$^{164}$Yb}  & 0.001&9.99&2.63&0.122 \\
				{$^{172}$Yb} & 0.61& 2.96&6.56& 0.48\\	
				{$^{174}$Yb} &0.66&3.037&7.51& 0.12 \\
				{$^{176}$Yb} &0.67& 2.60&5.31&0.29\\
				{$^{176}$Hf}& 0.001&9.99& 5.79&0.24 \\	
				{$^{178}$Hf} & 0.19& 6.76& 4.79&0.25\\
				{$^{180}$Hf} & 0.001& 1.24& 4.36&0.99 \\
				{$^{182}$W}& 0.001&9.95&4.30& 0.92 \\
				{$^{184}$W}& 0.001& 9.98&2.73& 0.95\\
				{$^{186}$W}& 0.001& 3.72&1.90& 0.99\\
				{$^{188}$Os}& 0.001& 9.99&1.49& 0.32 \\
				{$^{230}$Th} & 0.001&9.99&5.22& 0.33 \\		
				\hline
	\end{tabular}}}
\end{table*}

\begin{table*}[tbph]
	\centering
	{\renewcommand{\arraystretch}{1}
		{\setlength{\tabcolsep}{1cm} 
			\caption{Comparison of experimental data (upper line) for energy spectra of E(5) to theoretical predictions (lower line) by the Bohr Hamiltonian with the P.T potential, Davidson potential (D)\cite{Bonatsos2} and Kratzer potential(K)\cite{Bonatsos3}.
			}
			\label{Table3}
			\begin{tabular}{|cccccc|}		
				\hline 
				Nucleus &&$ {R_{4,2}}$ & $ {R_{0,2}}$&$ {R_{2,2}}$&$\sigma_{r.m.s.}$\\
				\hline	
				{$^{98}$Ru} &{\scriptsize Exp}&2.14&2.0&2.2 &\\
				&	{\scriptsize D}&2.14&2.4&2.1&0.27 \\
				&	{\scriptsize K}&2.34&2.0&2.3&0.81 \\
				&{\scriptsize PT}&2.14&2.56&2.14&0.61 \\
				{$^{102}$Pd}&{\scriptsize Exp}&2.29&2.9&2.8& \\
				&{\scriptsize D}& 2.24&2.3&2.2&0.32 \\
				&	{\scriptsize K}&2.42&2.4&2.4&0.99 \\
				&{\scriptsize PT}&2.20 &2.32&2.20&0.90 \\
				{$^{104}$Pd} &{\scriptsize Exp}&2.38 &2.4&2.4 &\\
				&{\scriptsize D}&2.21&2.6 &2.2&0.39 \\
				&	{\scriptsize K}&2.35&2.4&2.3&0.32 \\
				&{\scriptsize PT}&2.20&2.37 &2.20&0.86  \\
				{$^{106}$Pd} &{\scriptsize Exp}&2.40&2.2 &2.2& \\
				&{\scriptsize D}&2.16&2.2 &2.2&0.40 \\
				&	{\scriptsize K}&2.33&2.3&2.3&0.39 \\
				&{\scriptsize PT}&2.16&2.38 &2.16&0.96  \\
				{$^{108}$Pd}&{\scriptsize Exp}&2.42&2.4 &2.1 &\\
				&{\scriptsize D}&2.26&2.3 &2.3&0.31 \\
				&	{\scriptsize K}&2.38&2.5&2.4&0.31 \\
				&{\scriptsize PT}&2.23& 2.23&2.23&0.81 \\
				{$^{112}$Pd}&{\scriptsize Exp}&2.53 &2.6&2.1 &\\
				&{\scriptsize D}& 2.29&2.5&2.3&0.48 \\
				&	{\scriptsize K}&2.32&2.6&2.3&0.48 \\
				&{\scriptsize PT}&2.35&2.74 &2.35&0.96  \\
				{$^{114}$Pd} &{\scriptsize Exp}&2.56 &2.6&2.1& \\
				&{\scriptsize D}&2.31&2.8 &2.3&0.72 \\
				&	{\scriptsize K}&2.40&2.6&2.4&0.77 \\
				&{\scriptsize PT}&2.36&2.81 &2.36&0.44  \\
				{$^{106}$Cd}&{\scriptsize Exp}&2.36&2.8&2.7& \\
				&{\scriptsize D}&2.25&2.9&2.3&0.26 \\
				&	{\scriptsize K}&2.33&2.8&2.3&0.17 \\
				&{\scriptsize PT}&2.23&2.94&2.23& 0.72 \\
				{$^{108}$Cd}  &{\scriptsize Exp}&2.38&2.7&2.52& \\
				&{\scriptsize D}&2.14&2.2&2.1&0.5 \\
				&	{\scriptsize K}&2.29&1.9&2.3&0.34 \\
				&{\scriptsize PT}&2.16&2.54&2.16&0.92  \\
				{$^{110}$Cd} &{\scriptsize Exp}&2.35&2.2&2.2& \\
				&{\scriptsize D}&2.08&1.9&2.1 &0.41\\
				&	{\scriptsize K}&2.29&1.9&2.3&0.34 \\
				&{\scriptsize PT}&2.08&2.06&2.08&0.98  \\
				{$^{114}$Cd} &{\scriptsize Exp}&2.30&2.0&2.2& \\
				&{\scriptsize D}&2.06&1.9&2.1 &0.41\\
				&	{\scriptsize K}&2.25&1.7&2.2 &0.24\\
				&{\scriptsize PT}&2.05&2.04&2.05& 0.84 \\
				{$^{116}$Cd}  &{\scriptsize Exp}&2.38&2.5&2.4&\\
				&{\scriptsize D}&2.16&2.7&2.2&0.38  \\
				&	{\scriptsize K}&2.27&2.8&2.3&0.30 \\
				&{\scriptsize PT}&2.16&2.69&2.16&0.92  \\
				{$^{120}$Cd} &{\scriptsize Exp}&2.38&2.7& 2.6&\\
				&{\scriptsize D}&2.20&2.9&2.2&0.41\\
				&	{\scriptsize K}&2.31&2.7&2.3&0.42 \\
				&{\scriptsize PT}&2.20&2.95&2.20&0.68  \\
				
				\hline 
	\end{tabular}}}
\end{table*}

\begin{table*}[tbph]
	\centering
	{\renewcommand{\arraystretch}{1}
		{\setlength{\tabcolsep}{1cm} 
			\caption*{TABLE III. (Continued.)}
			\label{Table3a}
			\begin{tabular}{|cccccc|}		
				\hline 
				Nucleus &&$ {R_{4,2}}$ & $ {R_{0,2}}$&$ {R_{2,2}}$&$\sigma_{r.m.s.}$\\
				\hline
				{$^{128}$Xe} &{\scriptsize Exp}&2.33&3.6&2.2& \\
				&{\scriptsize D}&2.27&3.5&2.3&0.43 \\
				&	{\scriptsize K}&2.31&3.7&2.3&0.45 \\
				&{\scriptsize PT}&2.37&3.50&2.37& 0.89 \\
				{$^{130}$Xe}&{\scriptsize Exp}&2.25&3.3&2.1&\\
				&{\scriptsize D}&2.21&3.1&2.2&0.34  \\
				&	{\scriptsize K}&2.30&3.3&2.3&0.47 \\
				&{\scriptsize PT}&2.23&3.19&2.23&0.63  \\	
				{$^{132}$Xe} &{\scriptsize Exp}&2.16&2.8&1.9& \\
				&{\scriptsize D}&2.00&2.0&2.0&0.4 \\
				&	{\scriptsize K}&2.03&2.0&2.0&0.37 \\
				&{\scriptsize PT}&2.18&2.81&2.18&0.75  \\
				{$^{134}$Xe}&{\scriptsize Exp}&2.04&1.9&1.9& \\
				&{\scriptsize D}&2.00&2.0&2.0&0.68 \\
				&	{\scriptsize K}&1.87&1.6&1.9&0.21 \\
				&{\scriptsize PT}&2.03&2.06&2.03& 0.56 \\
				
				{$^{132}$Ba} &{\scriptsize Exp}&2.43&3.2&2.2& \\
				&{\scriptsize D}&2.29&2.8&2.3&0.61 \\
				&	{\scriptsize K}&2.37&3.2&2.4&0.76 \\
				&{\scriptsize PT}&2.34&3.70&2.34&0.97  \\
				
				{$^{134}$Ba}&{\scriptsize Exp}&2.32&2.9&1.9& \\
				&{\scriptsize D}&2.16&2.7&2.2&0.33 \\
				&	{\scriptsize K}&2.20&2.8&2.2&0.34 \\
				&{\scriptsize PT}&2.03&2.06&2.03&0.56  \\
				
				{$^{136}$Ba} &{\scriptsize Exp}&2.28&1.9&1.9& \\
				&{\scriptsize D}&2.00&2.0&2.0&0.25 \\
				&	{\scriptsize K}&2.00&1.9&2.0&0.19 \\
				&{\scriptsize PT}&2.03&2.05&2.03&0.58  \\
				{$^{136}$Ce} &{\scriptsize Exp}&2.38&1.9&2.0& \\
				&{\scriptsize D}&2.11&2.1&2.1& 0.45\\
				&	{\scriptsize K}&2.28&1.9&2.3&0.54 \\
				&{\scriptsize PT}&2.07&2.09&2.07&0.97 \\
				{$^{142}$Gd}&{\scriptsize Exp}&2.35&2.7&1.9& \\
				&{\scriptsize D}&2.21&2.8&2.2&0.23 \\
				&	{\scriptsize K}&2.33&2.6&2.3& 0.29\\
				&{\scriptsize PT}&2.20&2.86&2.20&0.38  \\
				{$^{144}$Gd} &{\scriptsize Exp}&2.35&2.5&2.5&\\
				&{\scriptsize D}&2.33&2.5&2.3&  0.12\\
				&	{\scriptsize K}&2.35&2.5&2.3&0.10 \\
				&{\scriptsize PT}&2.29&2.54&2.29&0.88  \\
				
				{$^{194}$Pt} &{\scriptsize Exp}&2.47&3.9&1.9 &\\
				&{\scriptsize D}&2.36&3.6&2.4&0.66 \\
				&	{\scriptsize K}&2.39&3.9&2.4&0.68 \\
				&{\scriptsize PT}&2.39&3.32&2.39& 0.82 \\
				
				{$^{196}$Pt} &{\scriptsize Exp}&2.47&3.2&1.9 &\\
				&{\scriptsize D}&2.33&2.9&2.3&0.63 \\
				&	{\scriptsize K}&2.38&3.1&2.4&0.67 \\
				&{\scriptsize PT}&2.38&3.13&2.38&0.87  \\	
				
				{$^{198}$Pt}& {\scriptsize Exp}&2.42&2.2&1.9& \\
				&{\scriptsize D}&2.21&2.2&2.2&0.37 \\
				&	{\scriptsize K}&2.25&2.3&2.3&0.37 \\
				&{\scriptsize PT}&2.28&2.23&2.28&0.91  \\
				\hline 
	\end{tabular}}}	
\end{table*}

\begin{table*}[tbph]
	\centering
	{\renewcommand{\arraystretch}{1}
		{\setlength{\tabcolsep}{1cm} 
			\caption{Comparison of experimental data (upper line) for energy spectra of X(5) to theoretical predictions (lower line) by the Bohr Hamiltonian with the PT potential, Davidson potential (D)\cite{Bonatsos2} and Kratzer potential(K)\cite{Bonatsos3}.}
			\label{Table4}
			\begin{tabular}{|cccccc|}		
				\hline 
				Nucleus &&$ {R_{4,2}}$ & $ {R_{0,2}}$&$ {R_{2,2}}$&$\sigma_{r.m.s.}$\\
				\hline
				{$^{154}$Sm}  &{\scriptsize Exp}&3.25&13.4& 17.6 &\\
				&{\scriptsize D}& 3.27 &13.0&18.6&0.51\\
				&	{\scriptsize K}&3.29&13.6&18.6&0.50 \\
				&{\scriptsize PT}&3.27&13.71&18.51&0.70\\
				
				{$^{158}$Gd} &{\scriptsize Exp}&3.29&15.0&14.9\\
				&{\scriptsize D}&3.29&14.5&15.1&0.32\\
				&	{\scriptsize K}&3.30&14.8&15.2&0.14 \\
				&{\scriptsize PT}&3.28&14.31&15.14&0.43\\
				{$^{160}$Gd}&{\scriptsize Exp}&3.30&17.6&13.1&\\
				&{\scriptsize D}&3.30&17.3&13.2&0.12\\
				&	{\scriptsize K}&3.31&17.6&13.0&0.14 \\
				&{\scriptsize PT}&3.30&17.22&13.12&0.14\\
				{$^{162}$Gd}  &{\scriptsize Exp}&3.29&19.8&12.0\\
				&{\scriptsize D}&3.30&19.8&12.1&0.07\\
				&	{\scriptsize K}&3.31&19.9&11.9&0.09 \\
				&{\scriptsize PT}&3.30&19.85&12.05&0.03\\
				{$^{158}$Dy}&{\scriptsize Exp}&3.21&10.0&9.6&\\
				&{\scriptsize D}&3.22&9.6&10.3&0.49\\
				&	{\scriptsize K}&3.27&9.9&10.1&0.83 \\
				&{\scriptsize PT}&3.21&10.34&10.07&0.57\\
				{$^{162}$Dy} &{\scriptsize Exp}&3.29&17.3&11.0&\\
				&{\scriptsize D}&3.30&15.7&11.2&0.74\\
				&	{\scriptsize K}&3.30&15.5&11.0&0.83 \\
				&{\scriptsize PT}&3.29&15.41&11.14&0.85\\
				{$^{164}$Dy} &{\scriptsize Exp}&3.30&22.6&10.4&\\
				&{\scriptsize D}&3.30&22.5&10.2&0.1\\
				&	{\scriptsize K}&3.31&22.9&10.2&0.19 \\
				&{\scriptsize PT}&3.30&22.43&10.24&0.08\\
				{$^{166}$Dy} &{\scriptsize Exp}&3.31&15.0&11.2&\\
				&{\scriptsize D}&3.31&14.9&11.2&0.07\\
				&	{\scriptsize K}&3.30&15.0&11.2&0.06 \\
				&{\scriptsize PT}&3.30&14.81&11.19&0.11\\
				{$^{160}$Er}  &{\scriptsize Exp}&3.10&7.1&6.8&\\
				&{\scriptsize D}&3.16&8.1&6.6&0.69\\
				&	{\scriptsize K}&3.24&7.2&6.9&0.87 \\
				&{\scriptsize PT}&3.15&8.59&6.64&0.85\\	
				{$^{162}$Er} &{\scriptsize Exp}&3.23&10.7&8.8&\\
				&{\scriptsize D}&3.23&10.7&10.1&0.77\\
				&	{\scriptsize K}&3.27&10.6&9.7&0.51 \\
				&{\scriptsize PT}&3.22&10.75&10.11&0.80\\	
				{$^{166}$Er}  &{\scriptsize Exp}&3.39&18.1&9.8&\\
				&{\scriptsize D}&3.28&16.8&9.9&0.69\\
				&	{\scriptsize K}&3.29&17.6&10.0&0.34 \\
				&{\scriptsize PT}&3.28&16.84&10.03&0.70\\
				{$^{168}$Er}   &{\scriptsize Exp}&3.31&15.3&10.3&\\
				&{\scriptsize D}&3.31&14.4&10.2&0.40\\
				&	{\scriptsize K}&3.31&14.5&10.3&0.27 \\
				&{\scriptsize PT}&3.30&14.09&10.30&0.57\\

				\hline 
	\end{tabular}}}
\end{table*}
\begin{table*}[tbph]
	\centering
	{\renewcommand{\arraystretch}{1}
		{\setlength{\tabcolsep}{1cm} 
			\caption*{TABLE IV. (Continued.)}
			\label{Table4a}
			\begin{tabular}{|cccccc|}		
				\hline 
				Nucleus &&$ {R_{4,2}}$ & $ {R_{0,2}}$&$ {R_{2,2}}$&$\sigma_{r.m.s.}$\\
				\hline
				{$^{164}$Yb}  &{\scriptsize Exp}&3.13&7.9&7.0&\\
				&{\scriptsize D}&3.18&8.3&7.4&0.35\\
				&	{\scriptsize K}&3.24&7.9&7.2&0.77 \\
				&{\scriptsize PT}&3.18&8.94&7.27&0.45\\
				{$^{172}$Yb}  &{\scriptsize Exp}&3.31&13.2&18.6 &\\
				&{\scriptsize D}&3.30&12.2&18.9&0.72 \\
				&	{\scriptsize K}&3.31&12.7&18.8&0.78 \\
				&{\scriptsize PT}&3.30&14.20&18.85&0.70\\
				{$^{174}$Yb}&{\scriptsize Exp}&3.31&19.4&21.4 &\\
				&{\scriptsize D}&3.31&19.3&21.5&0.10 \\
				&	{\scriptsize K}&3.32&19.1&21.5&0.20 \\
				&{\scriptsize PT}&3.30&19.21&21.50&0.13 \\
				{$^{176}$Yb}&{\scriptsize Exp}&3.31&13.9&15.4 &\\
				&{\scriptsize D}&3.30&13.7&15.5&0.28 \\
				&	{\scriptsize K}&3.31&13.5&15.5&0.12 \\
				&{\scriptsize PT}&3.29&13.74&15.21&0.39\\
				{$^{178}$Hf}&{\scriptsize Exp}&3.29&12.9&12.6&\\
				&{\scriptsize D}&3.28&12.3&13.0&0.35\\
				&	{\scriptsize K}&3.29&12.9&13.0&0.14 \\
				&{\scriptsize PT}&3.26&12.29&13.35&0.49\\
				{$^{180}$Hf} &{\scriptsize Exp}&3.31&11.8&12.9&\\
				&{\scriptsize D}&3.30&11.5&13.0&0.15\\
				&	{\scriptsize K}&3.31&11.6&12.8&0.21 \\
				&{\scriptsize PT}&3.29&11.70&12.86&0.39\\	
				{$^{182}$W} &{\scriptsize Exp}&3.29&11.3&12.2 &\\
				&{\scriptsize D}&3.29&11.5&12.5&0.19 \\
				&	{\scriptsize K}&3.31&11.5&12.4&0.18 \\
				&{\scriptsize PT}&3.27&11.52&12.38&0.35\\	
				{$^{184}$W} &{\scriptsize Exp}&3.27&9.0&8.1 &\\
				&{\scriptsize D}&3.28&8.9&8.0&0.09 \\
				&	{\scriptsize K}&3.29&9.1&8.1&0.09 \\
				&{\scriptsize PT}&3.24&9.55&7.96&0.53\\
				{$^{186}$W} &{\scriptsize Exp}&3.23&7.2&6.0& \\
				&{\scriptsize D}&3.25&7.2&6.3&0.13 \\
				&	{\scriptsize K}&3.29&7.5&6.1&0.15 \\
				&{\scriptsize PT}&3.23&8.43&5.71&0.80\\	
				{$^{188}$Os}&{\scriptsize Exp}&3.08&7.0&4.1 &\\
				&{\scriptsize D}&3.15&7.2&4.4&0.17 \\
				&	{\scriptsize K}&3.21&7.3&4.3&0.21 \\
				&{\scriptsize PT}&3.13&7.32&4.38&0.18 \\
				{$^{230}$Th}&{\scriptsize Exp}&3.27&11.9&14.7 &\\
				&{\scriptsize D}&3.27&11.6&14.7&0.24 \\
				&	{\scriptsize K}&3.30&11.7&14.7&0.62 \\
				&{\scriptsize PT}&3.26&12.24&14.45&0.44\\
				\hline 
	\end{tabular}}}
\end{table*}

{\renewcommand{\arraystretch}{1}
	\begin{table*}[tbph]
		\caption{Comparison of experimental data (upper line) for several B(E2) ratios of E(5) to predictions (lower line) by the Bohr Hamiltonian with the PT potential, Davidson potential (D)\cite{AIM2} and Kratzer potential(K)\cite{Bonatsos3}. We note in this case that our model (PT) has 3 free parameters, while models (Davidson and Kratzer ) possess only 2 free parameters.}
		\label{Table5}
		\begin{tabular}{ c c c c c c c c c c c}
			\hline  
			Noyaux&& $ \frac{4_{g}\rightarrow2_{g}}{2_{g}\rightarrow0_{g}} $ & $\frac{6_{g}\rightarrow4_{g}}{2_{g}\rightarrow0_{g}} $ & 
			$\frac{8_{g}\rightarrow6_{g}}{2_{g}\rightarrow0_{g}} $ & $\frac{10_{g}\rightarrow8_{g}}{2_{g}\rightarrow0_{g}} $ &  $\frac{2_{\gamma}\rightarrow2_{g}}{2_{g}\rightarrow0_{g}}  $ & $\frac{2_{\gamma}\rightarrow0_{g}}{2_{g}\rightarrow0_{g}}$$\times$$10^{3} $& $\frac{0_{\beta}\rightarrow4_{g}}{2_{g}\rightarrow0_{g}}  $ & 
			$\frac{2_{\beta}\rightarrow0_{\beta}}{0_{g}\rightarrow0_{g}}  $$\times$$10^{3} $ &$\sigma_{r.m.s.}$ \\
			
			\hline
			{$^{98}$Ru} &{\scriptsize  Exp}&1.44(25)&&&&1.62(61) &36.0(152)&&	\\
			&{\scriptsize D}&1.82& 2.62&	3.42 &4.22&	1.82& 0.0&1.36&3.60&0.14\\
			&	{\scriptsize K}&1.77&2.81&4.63&8.42&1.77 &0.0&1.27&	27.84&0.11 \\
			&{\scriptsize PT}&1.75&2.32& 2.85&3.34&1.75&0.&1.15&2.44&0.11\\
			
			{$^{102}$Pd} &{\scriptsize Exp}&1.56(19)&&&&0.46(9)& 128.8(735)&&\\
			&{\scriptsize D}&	1.76 &2.49&3.19 &3.87&1.76& 0.0&1.22&12.34&0.43\\
			&	{\scriptsize K}&1.63&2.31&3.25&4.77&1.63& 0.0 
			&0.87&41.64&0.39\\
			&{\scriptsize PT}&1.67& 2.06& 2.35&2.57&1.67& 0.& 0.86& 0.07&0.40\\
			
			{$^{104}$Pd} &{\scriptsize Exp}&1.36(27)&&&&0.61(8)& 33.3(74)&&\\
			&{\scriptsize D}&	1.74 &2.45&	3.15 &3.85&1.74 &0.0&1.11&8.13&0.39\\
			&	{\scriptsize K}&1.70&2.52&3.74&5.83&1.70 &0.0&0.99&	24.16&0.37\\
			&{\scriptsize PT}&1.67& 2.07&2.36& 2.60& 1.67& 0.& 0.86& 0.19&0.36\\
			
			{$^{106}$Pd}&{\scriptsize Exp}&1.63(28)&&&&0.98(12) &26.2(31)& 0.67(18)&&\\
			&{\scriptsize D}&	1.85& 2.67&3.49& 4.28 &1.85& 0.0&1.49 
			&5.98&0.30\\
			&	{\scriptsize K}&1.74 &
			2.66&4.13&	6.83&1.74 &0.0&	1.12& 22.91& 0.22\\
			&{\scriptsize PT}&1.72& 2.20& 2.59&2.91&1.72& 0.&1.05& 0.53&0.20\\
			
			{$^{108}$Pd} &{\scriptsize Exp}&1.47(20)&2.16(28)&
			2.99(48)&&1.43(14)& 16.6(18)&	1.05(13) &1.09(29) &\\
			&{\scriptsize D}&1.75 &2.45&3.12 &3.75&	1.75 &0.0&	1.20&15.82&0.07\\
			&	{\scriptsize K}&	1.66&2.38&3.42&5.11&	0.89&1.66 &0.0 	&30.31& 0.08\\
			&{\scriptsize PT}&1.65&1.99& 2.22&2.40& 1.65& 0.& 0.77& 0.007&0.12\\
			
			{$^{106}$Cd}&{\scriptsize Exp}&1.78(25)&&&&0.43(12)& 93.0(127)&&\\
			&{\scriptsize D}&	1.68& 2.32&2.95& 3.58&1.68& 0.0&0.92&10.44&0.41\\
			&	{\scriptsize K}&1.66 &2.37& 3.34 &4.76&1.66& 0.0&	0.83&16.97 &0.41\\
			&{\scriptsize PT}&1.65&2.08&2.44&2.77&1.65& 0.& 0.78&3.80&0.41\\
			
			{$^{108}$Cd} &{\scriptsize Exp}&1.54(24)&&&&0.64(20) &67.7(120)&&\\
			&{\scriptsize D}&	1.85 &2.69&3.52 &4.35&	1.85 &0.0&	1.49&4.06 &0.41\\
			&	{\scriptsize K}&
			1.67& 2.40& 3.43& 5.01&1.67& 0.0 &
			0.87&19.88&0.3460
			\\
			&{\scriptsize PT}&1.73&2.25&2.71& 3.12&1.73& 0.&1.07& 1.84&0.36\\	
			
			{$^{110}$Cd}&{\scriptsize Exp}&1.68(24)&&&&1.09(19)& 48.9(78)&&9.85(595) \\
			&{\scriptsize D}&	1.99& 2.97 &	3.93& 4.87  &1.99& 0.0  & 	1.98 & 1.61	&0.23\\
			&	{\scriptsize K}&1.85 &3.14& 5.63 &11.54&1.85& 0.0&1.52&20.99&0.19\\
			&{\scriptsize PT}&1.84& 2.43&2.91& 3.31&1.84& 0.& 1.47& 0.01&0.19\\
			
			{$^{114}$Cd} &{\scriptsize Exp}&1.99(25)  & 3.83(72) & 2.73(97) &&0.71(24)& 15.4(29) & 0.88(11) &10.61(193)& \\
			&{\scriptsize D}&	2.00& 2.99 	&3.97 &4.94&  	2.00& 0.0 &1.99 &0.74&0.32\\
			&	{\scriptsize K}&1.93& 3.46& 6.72 &15.44&	1.93 &0.0&1.77&15.44&0.61\\
			&{\scriptsize PT}&1.90&2.56&3.14&3.65&1.90&0.&1.65& 0.004&0.27\\
			
			{$^{116}$Cd}&{\scriptsize  Exp}&1.70(52)&&&&0.63(46)& 32.8(86)& 0.02&&\\
			&{\scriptsize D}& 1.74& 2.46  &	3.17 &3.90 &	1.74 &0.0 	& 1.11 	&4.42&0.38	\\
			&	{\scriptsize K}&1.69 &2.47 &3.52 &5.05&	1.69 &0.0&0.90 	&10.02&  0.34 \\
			&{\scriptsize PT}&1.74&2.32& 2.86&3.40&1.74&0.&1.12& 3.87&0.39\\
			
			{$^{128}$Xe}&{\scriptsize  Exp}&1.47(20)&1.94(26) &
			2.39(40) &2.74(114)&1.19(19)& 15.9(23) & \\
			&{\scriptsize D}&1.63& 2.20&2.75& 3.31&1.63& 0.0 &	0.73 &9.64&0.14\\
			&	{\scriptsize K}&7.64 &1.83 &0.0& 0.75 &	1.83& 2.95 &4.73&12.57 &0.92\\
			&{\scriptsize PT}&1.52& 1.77& 1.95& 2.09&1.52& 0.& 0.32& 2.79&0.14\\
			
			{$^{132}$Xe} &{\scriptsize  Exp}&1.24(18)&&&&1.77(29) &3.4(7) &&& \\
			&{\scriptsize D}&2.00 &3.00 &	4.00 &5.00 &2.00 &0.0& 2.00 &0.00 &0.26\\
			&	{\scriptsize K}&	2.78 &7.13& 17.89	&	43.35& 2.78& 0.0 &2.49 &0.07&0.61	\\
			&{\scriptsize PT}&1.71&2.23& 2.72&3.19& 1.71& 0.& 0.99&4.42&0.15\\

			\hline
		\end{tabular}
	\end{table*}
}

{\renewcommand{\arraystretch}{1}
	\begin{table*}[tbph]
		\caption*{TABLE V. (Continued.)}
		\label{Table5a}
		\begin{tabular}{ c c c c c c c c c c c}
			\hline  
			Noyaux&& $ \frac{4_{g}\rightarrow2_{g}}{2_{g}\rightarrow0_{g}} $ & $\frac{6_{g}\rightarrow4_{g}}{2_{g}\rightarrow0_{g}} $ & 
			$\frac{8_{g}\rightarrow6_{g}}{2_{g}\rightarrow0_{g}} $ & $\frac{10_{g}\rightarrow8_{g}}{2_{g}\rightarrow0_{g}} $ &  $\frac{2_{\gamma}\rightarrow2_{g}}{2_{g}\rightarrow0_{g}}  $ & $\frac{2_{\gamma}\rightarrow0_{g}}{2_{g}\rightarrow0_{g}}$$\times$$10^{3} $& $\frac{0_{\beta}\rightarrow4_{g}}{2_{g}\rightarrow0_{g}}  $ & 
			$\frac{2_{\beta}\rightarrow0_{\beta}}{0_{g}\rightarrow0_{g}}  $$\times$$10^{3} $ &$\sigma_{r.m.s.}$ \\
			
			\hline
			
			{$^{132}$Ba} &{\scriptsize Exp}&&&&&3.35(64)& 90.7(177)  &\\
			&{\scriptsize D}&1.68& 2.30 &2.90 &3.50 &1.68 &0.0 & 0.92&15.21&0.83\\
			&	{\scriptsize K}&1.61& 2.20 &2.94&3.95& 1.61& 0.0& 0.66 &20.59 &0.87	\\
			&{\scriptsize PT}&1.54& 1.85&2.08&2.28& 1.54& 0.&0.42& 5.41&0.90\\
			
			{$^{134}$Ba} &{\scriptsize Exp}&1.55(21)&&&&2.17(69)& 12.5(41) &&\\
			&{\scriptsize D}&1.75 &2.48 &	3.21& 3.94  &1.75 &0.0&1.14 & 4.08 &0.15	\\
			&	{\scriptsize K}&2.13 &4.10 &7.88&15.19 &2.13 & 0.0& 1.26&6.22&0.19\\
			&{\scriptsize PT}&1.93& 2.65& 3.31&3.93& 1.93& 0.& 1.75&0.01&0.14\\
			
			{$^{194}$Pt}&{\scriptsize Exp}&1.73(13)& 1.36(45)& 1.02(30) &0.69(19) &1.81(25) &5.9(9)& 0.01&& \\
			&{\scriptsize D}&1.59 &2.09 &2.57 &3.04& 1.59 &0.0 &0.63&19.78 &0.40\\
			&	{\scriptsize K}&1.56& 2.07 &2.63 &3.34&1.56 &0.0&0.52 &19.45&0.45\\
			&{\scriptsize PT}&1.50&1.72& 1.87&1.97& 1.50&0.& 0.25&1.44&0.23\\
			
			{$^{196}$Pt} &{\scriptsize Exp}&1.48(3) &1.80(23) &1.92(23) &&&0.4 &0.07(4)& 0.06(6)&\\
			&{\scriptsize D}&1.64 &2.21 &2.75 &3.28& 1.64 &0.0 &0.82&20.83&0.19 \\
			&	{\scriptsize K}&1.61 &2.21& 2.97 &4.04&1.61 &0.0&0.69 &23.11&0.21 \\
			&{\scriptsize PT}&1.50&1.73& 1.87&1.98&1.50&0.&0.27&1.02&0.03\\
			
			{$^{198}$Pt} &{\scriptsize Exp}&1.19(13)& $>$1.78&&&1.16(23)& 1.2(4) &0.81(22) &1.56(126)&\\
			&{\scriptsize D}&1.82 &2.60 &3.36&4.08 &1.82 &0.0 &1.41&10.09&0.22\\
			&	{\scriptsize K}&1.76 &2.73 &4.24 &6.76&1.76 &0.0&1.16 &11.09&0.21\\
			&{\scriptsize PT}&1.59& 1.87&2.05&2.17&1.59&0.& 0.58& 0.05&0.10\\
			\hline 
		\end{tabular}
\end{table*}}
{\renewcommand{\arraystretch}{0.5} 
	\begin{table*}[tbph]
		\caption{Comparison of experimental data (upper line) for several B(E2) ratios of X(5) to predictions (lower line) by the Bohr Hamiltonian with the PT potential, Davidson potential (D)\cite{AIM2} and Kratzer potential(K)\cite{Bonatsos3}. We note in this case that our model (PT) has 4 free parameters, while models (Davidson and Kratzer ) possess only 3 free parameters.}
		\label{Table6}
		\begin{tabular}{ c c c ccccc c c c c c c c }
			\hline 
			Noyaux&& $ \frac{4_{g}\rightarrow2_{g}}{2_{g}\rightarrow0_{g}} $ & $\frac{6_{g}\rightarrow4_{g}}{2_{g}\rightarrow0_{g}} $ & 
			$\frac{8_{g}\rightarrow6_{g}}{2_{g}\rightarrow0_{g}} $ & $\frac{10_{g}\rightarrow8_{g}}{2_{g}\rightarrow0_{g}} $ &  $\frac{2_{\beta}\rightarrow0_{g}}{2_{g}\rightarrow0_{g}}  $ & $\frac{2_{\beta}\rightarrow2_{g}}{2_{g}\rightarrow0_{g}}$& $\frac{2_{\beta}\rightarrow4_{g}}{2_{g}\rightarrow0_{g}}  $& 
			$\frac{2_{\gamma}\rightarrow0_{g}}{0_{g}\rightarrow0_{g}}  $&	$\frac{2_{\gamma}\rightarrow2_{g}}{0_{g}\rightarrow0_{g}}  $& $\frac{2_{\gamma}\rightarrow4_{g}}{0_{g}\rightarrow0_{g}}  $ &$\sigma_{r.m.s.}$&\\
			&&&&&$\times$$10^{3} $&$\times$$10^{3} $&$\times$$10^{3} $&$\times$$10^{3} $&$\times$$10^{3} $&$\times$$10^{3} $&&\\
			\hline	
			{$^{154}$Sm} &{\scriptsize Exp}&1.40(5) &1.67(7) &1.83(11)& 1.81(11)&
			5.4(13)&& 25(6) &18.4(34) &&
			3.9(7) \\
			&{\scriptsize D}&1.47 &1.69& 1.87 &2.06&26.7& 50.0 &150 &47.5&69.6& 3.7&0.03	\\
			&	{\scriptsize K}&1.46 &1.67 &1.86 &2.05 &24.7 &45.7&	136& 47.8 &69.9 &3.7&0.03 \\
			&{\scriptsize PT}&1.45& 1.66&1.82&1.96& 0.57&1.14&3.72&46.56& 68.13&3.57&0.02\\
			
			{$^{158}$Gd} &{\scriptsize Exp}& 1.46(5)&& 1.67(16) &1.72(16)&1.6(2)& 0.4(1)&7.0(8)&17.2(20)&30.3(45) &1.4(2) \\
			&{\scriptsize D}&1.46& 1.66 &1.82 &1.98&	25.7 &45.9&127& 64.0&93.0 &4.8&0.03	\\
			&	{\scriptsize K}&1.47 &1.69 &1.90& 2.13 &30.8 &56.5&	166& 70.7 &103.3& 5.4&0.05\\
			&{\scriptsize PT}&1.45& 1.64& 1.78&1.90& 0.66& 1.30&4.17& 61.64&89.49&4.62&0.02\\
			
			{$^{158}$Dy}  &{\scriptsize Exp}&1.45(10)& 1.86(12) &1.86(38)& 1.75(28)&12(3)& 19(4) &66(16) &32.2(78)&	103.8(258) &11.5(48) \\
			&{\scriptsize D}&1.50 &1.78 &2.04 &2.31&	30.5& 65.4&232& 88.5 &	131.7 &7.1&0.06\\
			&	{\scriptsize K}&1.48 &1.73 &1.98& 2.28& 32.5 &63.0&	202 &97.9 &
			143.6 &7.6&0.05 \\
			&{\scriptsize PT}&1.48& 1.73& 1.94&2.16&0.67&1.51&	5.71& 89.35& 132.28&7.06&0.04\\	
			
			{$^{162}$Dy} &{\scriptsize Exp}&1.45(7)& 1.51(10) &1.74(10) &1.76(13) &&&&0.12(1) &0.20 &0.02 \\
			&{\scriptsize D}&1.45 &1.65& 1.80 &1.95& 23.9 &42.4 &116 &89.8 &129.8& 6.7&0.04\\
			&	{\scriptsize K}&	1.45 &1.65 &1.80 &1.95 &23.7 &41.4&112& 92.3&133.2&6.8&0.04	\\
			&{\scriptsize PT}&1.44& 1.63&1.76&1.88& 0.58& 1.13&3.49& 88.34&127.65&6.52&0.03\\
			
			{$^{164}$Dy} &{\scriptsize Exp}&1.30(7) &1.56(7) &1.48(9) &1.69(9) &&&&19.1(22) &38.3(39) &4.6(5) \\
			&{\scriptsize D}&1.44& 1.62 &1.75 &1.86 &16.9& 29.1 &77 &99.7&	143.4& 7.3&0.05	\\
			&	{\scriptsize K}&1.45& 1.64 &1.79 &1.93&23.6& 40.6&107 &100.4&	144.7& 7.4& 0.06 \\
			&{\scriptsize PT}&1.44& 1.62&1.74&1.85& 0.01& 0.08&	0.07& 98.81&142.12&7.20&0.05\\
			
			{$^{162}$Er}&{\scriptsize Exp}&&&&&8(7) &&170(90)& 32.5(28) &77.0(56) &9.4(69) \\
			&{\scriptsize D}&1.49 &1.75 &1.99 &2.24& 27.8& 58.3& 202 &91.1&134.8 &7.2&0.01	\\
			&	{\scriptsize K}&1.48&	1.73&1.97&2.25& 28.9 &57.1 &189&100.4& 147.1 &7.8&0.02
			\\
			&{\scriptsize PT}&1.48& 1.72&1.94&2.15& 0.54&1.20&4.43& 89.70&132.57&7.06&0.03\\
			
			{$^{166}$Er} &{\scriptsize Exp}&1.45(12)& 1.62(22) &1.71(25) &1.73(23) &&&&25.7(31) &45.3(54)&3.1(4) \\
			&{\scriptsize D}&1.46 &1.66& 1.81 &1.96 &20.7 &38.2 &111 &100.0 &	144.8 &7.4&0.04	\\
			&	{\scriptsize K}&1.48 &1.74 &2.00 &2.31 &
			21.2 &39.2 &117&&
			\\
			&{\scriptsize PT}&1.45&1.65&1.81&1.96& 0.02&0.05&0.15& 98.64&142.80&7.33&0.04\\	
			
			{$^{168}$Er} &{\scriptsize Exp}&1.54(7) &2.13(16) &1.69(11) &1.46(11) &&&&23.2(15) &41.1(31) &3.0(3) \\
			&{\scriptsize D}&1.45 &1.65 &1.79 &1.93 &27.7 &47.2& 120 &100.6&	145.1 &7.4&0.09\\
			&	{\scriptsize K}&1.45&1.64&1.78&1.92 &27.6& 46.2 &116&100.6& 144.9 &7.4
			&	0.09\\
			&{\scriptsize PT}&1.44& 1.61&1.72&1.81&0.67&1.35&4.38&97.25&140.05&	7.11&0.09\\
			
			{$^{172}$Yb} &{\scriptsize Exp}&1.42(10)& 1.51(14) &1.89(19) &1.77(11) &1.1(1) &3.7(6) &12(1)& 6.3(6) &&0.6(1) \\
			&{\scriptsize D}&1.46 &1.67 &1.83 &1.99& 32.2& 55.9 &147 &51.6 &75.0 &3.9&0.03\\
			&	{\scriptsize K}&1.46&1.66&1.83&	2.01& 29.6 &51.6 &139&51.0 &74.2 &3.8&0.03\\
			&{\scriptsize PT}&1.44&1.61&1.72&1.81&0.68& 1.35&4.37& 49.93&72.14&3.69&0.02\\
			
			{$^{174}$Yb}&{\scriptsize Exp}&1.39(7) &1.84(26) &1.93(12) &1.67(12)&&&&&12.4(29)&	\\
			&{\scriptsize D}&1.45 &1.63 &1.75 &1.86 &20.9& 35.1& 88& 45.0&64.9& 3.3&0.06\\
			&	{\scriptsize K}&1.44&1.62&1.74&1.86 &20.6& 34.3& 85&45.7 &65.8 &3.4&
			0.07\\
			&{\scriptsize PT}&1.44& 1.61&1.72&	1.81& 0.43& 0.77&2.18&44.01& 63.44&3.23&0.06\\
			
			{$^{176}$Yb}&{\scriptsize Exp}&1.49(15)& 1.63(14) &1.65(28) &1.76(18) &&&&9.8 &\\
			&{\scriptsize D}&1.46& 1.66 &1.82& 1.97& 27.9& 49.0 &132 &63.1&91.6 &4.7 &0.05\\
			&	{\scriptsize K}&1.45&1.65&1.81&1.97&28.6 &49.0& 128&64.5& 93.4& 4.8&
			0.05\\
			&{\scriptsize PT}&1.44& 1.63& 1.75&1.85& 0.75&1.51&4.95&62.25& 90.15&4.63&0.03\\

			\hline 
		\end{tabular}
		
	\end{table*}
}

{\renewcommand{\arraystretch}{0.5} 
	\begin{table*}[tbph]
		\caption*{TABLE VI. (Continued.)}
		\label{Table6a}
		\begin{tabular}{ c c c ccccc c c c c c c c }
			\hline 
			Noyaux&& $ \frac{4_{g}\rightarrow2_{g}}{2_{g}\rightarrow0_{g}} $ & $\frac{6_{g}\rightarrow4_{g}}{2_{g}\rightarrow0_{g}} $ & 
			$\frac{8_{g}\rightarrow6_{g}}{2_{g}\rightarrow0_{g}} $ & $\frac{10_{g}\rightarrow8_{g}}{2_{g}\rightarrow0_{g}} $ &  $\frac{2_{\beta}\rightarrow0_{g}}{2_{g}\rightarrow0_{g}}  $ & $\frac{2_{\beta}\rightarrow2_{g}}{2_{g}\rightarrow0_{g}}$& $\frac{2_{\beta}\rightarrow4_{g}}{2_{g}\rightarrow0_{g}}  $& 
			$\frac{2_{\gamma}\rightarrow0_{g}}{0_{g}\rightarrow0_{g}}  $&	$\frac{2_{\gamma}\rightarrow2_{g}}{0_{g}\rightarrow0_{g}}  $& $\frac{2_{\gamma}\rightarrow4_{g}}{0_{g}\rightarrow0_{g}}  $ &$\sigma_{r.m.s.}$&\\
			&&&&&$\times$$10^{3} $&$\times$$10^{3} $&$\times$$10^{3} $&$\times$$10^{3} $&$\times$$10^{3} $&$\times$$10^{3} $&&\\
			\hline	
			
			{$^{176}$Hf} &{\scriptsize Exp}&&&&&5.4(11) &&31(6) &21.3(26) &&\\
			&{\scriptsize D}&1.47& 1.70 &1.89& 2.09 &30.8& 57.3& 169& 57.9&84.9 &4.5	&0.04\\
			&	{\scriptsize K}&	1.46& 1.68 &1.86 &2.06 &29.1 &52.2& 148 &60.3&87.8&
			4.6&0.04\\
			&{\scriptsize PT}&1.45&1.66&1.82&1.96&0.74&1.53&	5.20& 56.11&82.06&4.29&0.01\\
			
			{$^{178}$Hf}&{\scriptsize Exp}&&1.38(9) &1.49(6) &1.62(7) &0.4(2) &&2.4(9) &24.5(39)&27.7(28)& 1.6(2) \\
			&{\scriptsize D}&1.47& 1.69 &1.88 &2.07&	28.4& 53.1 &158 &73.8 &107.8 &5.6&0.08	\\
			&	{\scriptsize K}&1.46 &1.68& 1.86 &2.06 &27.1 &49.2 &142 &75.7&110.0&5.7&0.08 \\
			&{\scriptsize PT}&1.45&1.66&1.82&1.96&0.82&1.72&5.94&68.804&100.45&5.24&0.06\\	
			
			{$^{180}$Hf}&{\scriptsize Exp}&1.48(20) &1.41(15) &1.61(26) &1.55(10)&&&&	24.5(47)&32.9(56)&\\
			&{\scriptsize D}&1.46& 1.66& 1.82 &1.98&34.9 &59.5& 151& 78.4 &	113.4 &5.8&0.09\\
			&	{\scriptsize K}&1.46 &1.66 &1.83 &2.00 &34.6& 58.6 &150 &80.3&116.0&6.0&0.09 \\
			&{\scriptsize PT}&1.44&1.61& 1.72&1.80&0.74&1.62&	5.75&75.99& 109.68&5.59&0.05\\
			
			{$^{182}$W}&{\scriptsize Exp}&1.43(8)& 1.46(16)& 1.53(14)& 1.48(14)&6.6(6) &4.6(6) &13(1) &24.8(12)&49.2(24) &0.2 \\
			&{\scriptsize D}&1.47& 1.69 &1.87 &2.04 &32.5& 58.3& 162 &79.9 &116.2& 6.0&0.07\\
			&	{\scriptsize K}&1.46 &1.67 &1.85 &2.05 &33.0& 57.5 &155 &82.0&118.9&6.1&0.07
			\\
			&{\scriptsize PT}&1.45& 1.64&1.77&1.88& 0.95&2.074&7.42& 76.82&111.59&5.75&0.05\\
			
			{$^{184}$W} &{\scriptsize Exp}&1.35(12) &1.54(9) &2.00(18) &2.45(51) &1.8(3)&&24(3) &37.1(28) &70.6(51)& 4.0(4) \\
			&{\scriptsize D}&1.48& 1.73 &1.95& 2.16& 40.7 &75.2 &216 &128.3& 187.3 &9.8&0.04	\\
			&	{\scriptsize K}&1.48& 1.73& 1.97 &2.27& 38.9& 71.8 &214 &128.4&187.1&9.8&0.04 \\
			&{\scriptsize PT}&1.46&1.67&1.82&1.96&1.22&2.95&11.71&122.19&178.52&9.27&0.06\\
			
			{$^{186}$W} &{\scriptsize Exp}&1.30(9)& 1.69(12) &1.60(12)& 1.36(36)&&&&41.7(92)& 91.0(201)&	\\
			&{\scriptsize D}&1.51& 1.80 &2.07& 2.34& 46.2& 91.9 &289 &165.7 &244.5 &12.9&0.18\\
			&	{\scriptsize K}&1.49 &1.77& 2.08 &2.48 &47.3& 89.1 &275& 174.0&254.4&13.3&0.20 \\
			&{\scriptsize PT}&1.46& 1.68&1.83&1.95& 1.35& 3.60&15.32& 176.14& 257.35&13.32&0.11\\
			
			{$^{188}$Os}&{\scriptsize Exp}&1.68(11) &1.75(11)& 2.04(15) &2.38(32)&&&&63.3(92) &202.5(304)& 43.0(74) \\
			&{\scriptsize D}&1.54 &1.89 &2.25 &2.63& 33.9 &83.9&344& 229.8 &345.2 &18.7&0.06\\
			&	{\scriptsize K}&1.52& 1.87 &2.29 &2.87& 33.6 &78.5 &330& 246.6&366.2&19.7&0.09 \\
			&{\scriptsize PT}&1.51& 1.81&2.08&2.33& 1.26& 3.64&16.87&227.37&340.13&18.21&0.04\\
			
			{$^{230}$Th}&{\scriptsize Exp}&1.36(8)&&&&	5.7(26) &&	20(11)&15.6(59)&28.1(100) &	1.8(11)  \\
			&{\scriptsize D}&1.47 &1.70 &1.90 &2.09& 30.0&56.4 &168&	63.6& 93.2 &4.9&0.03\\
			&	{\scriptsize K}&1.46 &1.68& 1.86 &2.07 &31.4 &55.6 &155& 66.9&	97.3& 
			5.1 
			&0.03\\
			&{\scriptsize PT}&1.45& 1.66& 1.81&1.95& 0.84&1.77&	6.13& 62.69& 91.58&4.78&0.02\\

			\hline 
		\end{tabular}
		
	\end{table*}
}


\begin{thebibliography}{}
	\bibitem{Iachello1} F. Iachello and A. Arima, {\it The Interacting Boson Model} (Cambridge University Press, Cambridge, 1987).
	\bibitem{bohr} A. Bohr, Mat. Fys. Medd. K. Dan. Vidensk. Selsk. 26, no. 14 (1952).
	F. Iachello, Phys. Rev. Lett. 91 (2003) 132502.
	\bibitem{Casten1} R.F. Casten, Nat. Phys. 2 (2006) 811.
	\bibitem{Cejnar1} P. Cejnar, J. Jolie, R., F. Casten, Rev. Modern Phys. 82 (2010) 2155.
	\bibitem{E5} F. Iachello, Phys. Rev. Lett. {\bf85}, 3580 (2000).
	\bibitem{X5} F. Iachello, Phys. Rev. Lett. {\bf87}, 052502 (2001).
	\bibitem{Hill} D. L. Hill and J. A. Wheeler, Phys. Rev., 89 :1102–1145, Mar 1953.
	\bibitem{Liu} X. Y. Liu, G. F. Wei, X. W. Cao, H. G. Bai, Int. J. Theor. Phys. 49 (2010) 343
    \bibitem{Hamzavi} M. Hamzavi and S. M. Ikhdari, Mol.Phys. 110 (2012) 3031
    \bibitem{Chabab} M. Chabab,  A. El Batoul, and M. Oulne J. Math. Phys. 56, 062111 (2015).
    \bibitem{AIM1} H. Ciftci, R. L. Hall, and N. Saad, J. Phys. A: Math. Gen. {\bf36}, 11807 (2003).
	\bibitem{AIM2} M. Chabab, A. Lahbas, and M. Oulne, Phys. Rev. C {\bf91}, 064307 (2015).
	\bibitem{AIM3} M. Chabab, A. Lahbas, M. Oulne, Eur. Phys. J. A 51 (2015) 131.
	\bibitem{AIM4} M. Chabab, A. El Batoul, M. Oulne, Z. Naturforsch. A 71 (1) (2016) 59–68.
	\bibitem{AIM5} M. Chabab, A. El Batoul, M. Hamzavi, et al., Eur. Phys. J. A 53 (2017) 157.
    \bibitem{AIM6} M. Chabab, A. El Batoul, A. Lahbas, M. Oulne, J. Phys. G: Nucl. Part. Phys. 43 (2016) 125107.
   \bibitem{AIM7} M. Chabab, A. El Batoul, A. Lahbas, M. Oulne,  Nucl. Phys. A 953 (2016) 158.
   \bibitem{AIM8} I. Inci, D. Bonatsos, and I. Boztosun, Phys. Rev. C. 84 (2011) 024309.
  \bibitem{AIM9} P. Buganu, M. Chabab, A. El Batoul, A. Lahbas, M. Oulne, Nucl. Phys. A970 (2018) 272–290.
 \bibitem{wilet} L. Wilets and M. Jean, Phys. Rev.  {\bf102}, 788 (1956).
 \bibitem{bes} D. R. B\'es, Nucl. Phys. {\bf10}, 373 (1959).
 \bibitem{rakavy} G. Rakavy, Nucl. Phys. {\bf4}, 289 (1957).
 \bibitem{davison}D. Bonatsos, E. A. McCutchan, N. Minkov, R. F. Casten, P. Yotov, D. Lenis, D. Petrellis, and I. Yigitoglu, Phys. Rev. C {\bf76}, 064312 (2007).
 \bibitem{IBudaca} A. I.Budaca and R. Budaca, Eur. Phys. J. Plus (2019) 134: 145.
\bibitem{Edmonds} A.R. Edmonds, Angular Momentum in Quantum Mechanics, Princeton University Press,    Princeton, NJ, 1957.
 \bibitem{Bonatsos} D. Bonatsos, D. Lenis, N. Minkov, P.P. Raychev, P.A. Terziev, Sequence of potentials interpolating between the u(5) and e(5) symmetries, Phys. Rev. C 69 (2004) 044316.
 \bibitem{bijker03} R. Bijker, R. Casten, V. Zamfir and A. McCutchan, Phys. Rev. C 68 (2003) 064304.
\bibitem{data}	http://www.nndc.bnl.gov/nndc/ensdf/.
\bibitem{Bonatsos2} D. Bonatsos, P.E. Georgoudis, D. Lenis, N. Minkov, C. Quesne, Phys. Rev. C 83 (2011) 044321 .
\bibitem{Bonatsos3} D. Bonatsos, P.E. Georgoudis, N. Minkov, D. Petrellis and C. Quesne, Phys. Rev. C 88 (2013) 034316.
\end{thebibliography}
\end{document}